# Hydrodynamic escape from the proto-lunar disk and the origin of the Earth-Moon volatile dichotomy

Kaveh Pahlevan[a,b] *, Andrew N. Youdin[c], Paolo A. Sossi[d]

a. Carl Sagan Center, SETI Institute, Mountain View, CA, USA
b. Earth and Planets Laboratory, Carnegie Institution for Science, Washington, DC, USA
c. University of Arizona, Steward Observatory and Lunar and Planetary Laboratory, Tucson, AZ, USA
d. Institute of Geochemistry and Petrology, Department of Earth and Planetary Sciences, ETH Zurich, Zurich, Switzerland

* Correspondence should be addressed to: kpahlevan@seti.org

Word count: 6500 words

References: 70

Figures: 4

Tables: 2

Revised for publication in *EPSL*

**Abstract**

Volatile elements – those that vaporize at low temperatures – are depleted in lunar rocks relative to terrestrial rocks. This systematic chemical depletion is evidence for vaporization and preferential removal of vapor from proto-lunar materials during the high-temperature processes accompanying lunar origin. Despite the robustness of these observations, the physical processes by which proto-lunar vapors were removed after the giant impact are not yet well-understood. Here, we show that toward the end of post-giant impact cooling history, Earth's atmosphere was dominated by carbon species (e.g., CO) and was spatially compact, behaving as a closed system retaining Earth's volatile inventory, whereas the proto-lunar disk atmosphere was dominated by H and $H_2$ and was spatially extended, developing into a hydrodynamic outflow analogous to the solar wind. We find that equilibrium $H_2$ recombination ($2H \rightarrow H_2$) in a partially-dissociated disk atmosphere produces a nearly isothermal structure, a feature known to activate outflows. The expected outflow was strong enough to propel proto-lunar volatiles from a Roche-interior ($r < 3R_E$) disk out of Earth's gravity field into interplanetary space in a cometary tail of gas and perhaps dust. The proposed model suggests that the dichotomy in volatile element abundances between the silicate Earth and Moon is a natural outcome of the hydrodynamics of magma ocean atmospheres and that elemental and isotopic abundances of endogenous lunar volatiles are diagnostic of gaseous outflows launched from the proto-lunar disk towards the end of its condensation.

## 1. Introduction

Among the earliest observations of the Apollo samples was the stark depletion of volatile elements in lunar rocks relative to terrestrial rocks (Ringwood and Kesson, 1977), a chemical signature pointing towards high-temperature open-system processes operating during lunar origin. These observations played an important role in the rise of the giant impact hypothesis (Cameron and Ward, 1976; Hartmann and Davis, 1975), which has since become widely accepted (Canup et al., 2023). The chemical depletion of volatile elements is accompanied by isotopic fractionation, for example in zinc, potassium, and rubidium (Nie and Dauphas, 2019; Paniello et al., 2012; Wang and Jacobsen, 2016), recording a mass-selective process associated with the loss. This process has been identified as evaporation into a slightly undersaturated medium at temperatures in excess of 3,000 K (Dauphas et al., 2022; Nie and Dauphas, 2019) or loss of equilibrium vapor at somewhat lower temperatures near 2,500 K (Wang et al., 2019). Despite the robust signatures of lunar volatile depletion, and the near-ubiquitous acceptance of the accretion of the Moon from a molten disk, the fluid dynamical processes by which volatile-rich vapors were removed from proto-lunar liquids in the energetic aftermath of the Moon-forming giant impact are not yet well-understood.

Recent assessments of post-giant impact outflows have concluded that hydrodynamic escape from the Earth's gravitational field was unlikely and that the proto-lunar disk atmosphere behaved as a closed system (Canup et al., 2023). Models of lunar volatile depletion have therefore focused on scenarios in which the lunar complement of volatiles were transported inwards in the proto-lunar disk and accreted by the Earth (Canup et al., 2015; Nie and Dauphas, 2019) or consideration of the hydrodynamics of escape from a proto-Moon near Earth's Roche limit compared to that isolated in space (Charnoz et al., 2021; Madeira et al., 2025; Tang and Young, 2020). However, the aforementioned models consider a lunar or proto-lunar disk atmosphere formed exclusively from evaporation of anhydrous silicate material which, at magmatic temperatures, is dominated by Na, K, and SiO, leading to high mean molecular weight atmospheres relatively impervious to escape. By contrast, geochemical evidence indicates that Earth's inventory of water was acquired before or during the Moon-forming giant impact (Fischer et al., 2024; Greenwood et al., 2018). The implication is that highly volatile species, such as $H_2$ or $H_2O$, were abundant in the proto-lunar disk atmosphere. The presence of highly volatile elements would lower the mean molecular weight of atmospheres, making them more susceptible to escape. Both the Earth and the proto-lunar disk are expected to have hosted volatile element atmospheres dominated by C-, O- and H-bearing species. Here, we show that towards the end of condensation history, the post-giant impact atmosphere of Earth was dominated by carbon species and was gravitationally bound, whereas the proto-lunar disk atmosphere was an inflated hydrogen-dominated structure that developed into a vigorous outflow of vaporized species out of Earth's gravity field and into interplanetary space.

Any model of hydrodynamic outflow from post-giant impact atmospheres involves assumptions about the chemical composition of the vaporized materials present. Prior assessments of proto-lunar disk outflows have adopted chemical models using simple oxides and silicates ($H_2O$, $SiO_2$, $Mg_2SiO_4$, $Fe_2SiO_4$) and their vaporization and dissociation products to characterize the gases present (Desch and Taylor, 2011; Genda and Abe, 2003; Nakajima and Stevenson, 2018; Pahlevan et al., 2016). The behavior of hydrogen – due to its abnormally low atomic mass – is a recurring topic in discussions of post-giant impact hydrodynamic escape (Desch and Taylor, 2011; Nakajima and Stevenson, 2018; Pahlevan et al., 2016; Thompson and Stevenson, 1988). Existing models assume any hydrogen present derives from the thermal dissociation of water vapor and is therefore accompanied by abundant oxygen. Here, we introduce the influence of Fe-rich metals, which scavenge oxygen, producing post-giant impact atmospheres that are oxygen-poor (Hirschmann, 2012). Under these conditions, $H_2$ and its dissociation product H become the dominant species in the proto-lunar disk atmosphere. We show that such partially-dissociated hydrogen-rich disk atmospheres are unable to attain hydrostatic balance and become unstable with respect to rapid flow out of the Earth's gravitational field and into interplanetary escape. By contrast, we show that the Earth's volatile atmosphere – by virtue of massive water dissolution in a deep terrestrial magma ocean – was dominated by carbon species (e.g., CO) and maintained a compact vertical structure with negligible thermal losses. We propose that a hydrodynamic outflow from the proto-lunar disk – but not the post-giant impact Earth – was responsible for the dichotomy in volatile element abundances between the silicate Earth and Moon first observed in Apollo samples. In contrast to the hottest post-giant impact thermal states in which both major (Pahlevan and Stevenson, 2007) and refractory (Lock et al., 2018; Pahlevan et al., 2011) elements can experience Earth-disk mixing and equilibration through a common silicate vapor atmosphere, the proposed outflows would have occurred towards the end of disk thermal history when silicate vapors had nearly fully condensed. We therefore consider that Earth-disk equilibration and hydrodynamic outflows could have operated sequentially and that these processes can be analyzed separately.

Two lines of evidence suggest that the proto-lunar disk atmosphere evolved via condensation and cooling into an oxygen-poor but hydrogen-rich atmosphere susceptible to outflow. First, there is compelling evidence for a lunar metallic core (Williams et al., 2014), such that the oxygen fugacity ($fO_2$) of the proto-lunar disk atmosphere was likely buffered by metal-silicate equilibria, as also occurred in magma oceans during Earth's accretion (Frost et al., 2008). $fO_2$ estimates during lunar core formation suggest ΔIW near or below -1 (Righter, 2002). Secondly, lunar basalts record similarly low oxygen fugacity as evidenced, for example, by the appearance in lunar rocks of metallic iron as a primary igneous phase (Reid et al., 1970). These measurements and estimates suggest that the material from which the Moon formed was characterized by low $fO_2$, 1-2 log units below the iron-wüstite (IW) buffer (Wadhwa, 2008). These values of $fO_2$ correspond to vapors characterized by $pH_2/pH_2O$ of 3-10 and $pCO/pCO_2$ of 9-30 at magmatic temperatures (Visscher and Fegley, 2013). Definition of the equilibrium speciation in a model proto-lunar disk atmosphere in the C-O-H system requires, in addition to temperature and $fO_2$, two constraints: the mass of H and the H/C ratio, both of which can be estimated from bulk silicate Earth (BSE). The partial pressures of these species are further fractionated, for a fixed bulk lunar volatile budget of H and C, via their differential solubilities in magma, which decline in the order $H_2O >> CO_2 > H_2 \sim CO$ (Hirschmann et al., 2012; Pan et al., 1991; Sossi et al., 2023; Yoshioka et al., 2019). Contrary to previous work, we predict water vapor is a minor species whereas molecular hydrogen is a major disk species. At the temperatures of interest, thermal dissociation is also significant. Equilibration of a vapor atmosphere with both Fe-rich metal and silicate at high temperature makes $H_2$ and its dissociation product H the dominant H-bearing species in the proto-lunar disk. Towards the end of its silicate and metallic condensation, the mean molecular weight of the increasingly hydrogen-rich disk atmosphere decreases, the scale height increases, and the atmosphere expands. We show below (in §3.3) that such an inflated structure crosses the threshold of hydrostatic breakdown and enters the regime of hydrodynamic outflows.

The outline of the paper is: §2 – summary of the model used to describe the composition, vertical structure, hydrodynamic stability, and outflow properties of C-O-H magma ocean atmospheres, §3

– results for the post-giant impact C-O-H atmosphere of the Earth and proto-lunar disk, §4 – discussion of the consequences for the origin and abundance of lunar volatiles, and §5 – summary and conclusions.

## 2. Model description

The Moon-forming giant impact is thought to produce a circumterrestrial disk with the release of enormous quantities of heat enveloping the entire planet-disk system in a silicate vapor atmosphere (Benz et al., 1987; Canup and Asphaug, 2001; Pahlevan and Stevenson, 2007; Thompson and Stevenson, 1988). Radiative cooling and silicate vapor condensation gradually leads these two systems to evolve into magmatic systems coexisting with atmospheres dominated by volatile elements. In this section, we describe an estimate of the volatile inventories (§2.1) and onset temperature (§2.2) relevant to such volatile element atmospheres on both the Earth and the proto-lunar disk. We then determine the chemical composition of such atmospheres in the C-O-H system by calculating equilibrium dissolution with metal-saturated, fully molten terrestrial and proto-lunar disk magma oceans and calculating molecular speciation using gas-phase equilibria, modulated by their solubilities in the silicate liquid phase (§2.3). With a given atmospheric composition, we determine vertical structure of volatile-element atmospheres using reactive multi-species adiabats (§2.4). Combining the molecular composition and temperature structure for these atmospheres, we assess stability with respect to hydrodynamic expansion into interplanetary space using criteria developed in the context of the solar wind (§2.5). For hydrodynamically outflowing atmospheres, we describe a method to estimate mass loss rates for comparison with the isotopic record (§2.6).

### 2.1. The terrestrial water and carbon abundance at the time of the Moon-forming giant impact

Because hydrogen is the gas propelling bulk volatile element outflows considered in this work, its terrestrial abundance at the time of the Moon-forming giant impact is an important parameter that determines the vigor of outflows and the entrained volatile element inventory (§3.4). Geochemical evidence indicates that Earth's inventory of water (or hydrogen) was mainly acquired before or during the Moon-forming giant impact (Fischer et al., 2024; Greenwood et al., 2018). Because the C-O-H atmosphere enveloping the terrestrial magma ocean was apparently a closed system (Fig. 3), the best estimate for the BSE hydrogen and carbon abundances at the time of the Moon-forming giant impact is the total abundance of these elements accreted to the BSE. We consider two budget items for hydrogen. First, the $H_2O$ abundance of the modern BSE has been estimated to be between 700-3000 ppm (Hirschmann, 2018; Marty, 2012). Second, up to an ocean-equivalent of hydrogen may have been lost from Earth during the Archean eon (Zahnle et al., 2019). Therefore, we conservatively adopt a post-giant impact BSE $H_2O$ abundance of 900 ppm or equivalently an H abundance of 100 ppm. For the BSE carbon content, we consider an H/C atomic ratio of 12 or equivalently an H/C mass ratio of 1 as inferred for modern BSE (Hirschmann and Dasgupta, 2009). H and C abundances in the proto-lunar disk are assumed to be identical with that of the post-giant impact Earth, because the disk is thought to inherit Earth's mantle composition during its formation either via direct disk injection (Canup, 2012) or due to system-wide turbulent mixing in the common silicate vapor atmosphere of the Earth and the proto-lunar disk, a process called turbulent mixing or equilibration (Pahlevan and Stevenson, 2007).

## 2.2. Onset temperatures of volatile element atmospheres

Radiative cooling of partially vaporized magma oceans produced by giant impacts leads to silicate condensation and ultimately generates atmospheres dominated by volatile elements. Immediately after the Moon-forming giant impact, temperatures of 4,000-8,000 K are commonly achieved in the nascent Earth-disk system, producing silicate vapor atmospheres with pressures of tens to hundreds of bars or more (Canup, 2004; Pahlevan et al., 2011; Visscher and Fegley, 2013). Earlier works have shown that the low molar abundance ($x_i$) of volatile species (e.g., H, $x_H$~$10^{-4}$) in a thick silicate vapor atmosphere hinders hydrodynamic escape, relegating any thermal escape to be diffusion-limited. However, owing to fast radiative cooling ($\sigma T^4$) at these temperatures, this rapidly-evolving system would have been short-lived and diffusion-limited escape would have been negligible on the timescales of interest (Nakajima and Stevenson, 2018; Pahlevan et al., 2016). Other processes – such as secular cooling and its associated silicate condensation – must have operated to separate the volatile elements and enable their selective removal from the proto-lunar system. Following the condensation of the major silicate (Fe, Mg, Si, O) and metallic (Fe, Ni) constituents on the post-giant impact Earth and proto-lunar disk, the atmospheres enveloping these reservoirs become dominated by exsolved volatile elements (mainly C and H) whose abundances and speciation depends on the solubilities and redox state of coexisting silicate melt (Gaillard et al., 2022). By strongly concentrating volatile elements into one reservoir, such atmospheres make possible efficient, highly selective hydrodynamic removal of volatile elements from co-existing magma ocean reservoirs. We approximate that both atmospheric composition and the mechanical status (hydrostatic versus hydrodynamic) of such atmospheres can be described in the model C-O-H system. We seek to identify the hottest thermal states at which C-O-H atmospheres prevail, because it is at the hottest thermal states that retention or thermally-driven loss of volatiles can be robustly evaluated. We define the onset temperature of C-O-H atmospheres as the temperature at which the sum of the pressures of volatile species ($P_{COH}$) exceeds the silicate vapor pressure ($P_{SV}$), the latter being strongly dependent on temperature. For the proto-lunar disk, although C-O-H volatiles are extensively outgassed, low gravity and low surface density results in low volatile element pressure $P_{COH}$≈0.04 bars (Table 1 and Figure 2), which only predominates over silicate vapor pressure ($P_{SV}$) below $T_{COH}$≈2,800 K (see Appendix A for a discussion of the factors that can shift this estimate and produce the adopted range $T_{COH}$=2,800-3,200 K). For the post-giant impact Earth, a stronger gravitational acceleration and a larger surface density of the atmosphere results in a much higher pressure for the exsolved C-O-H atmosphere of $P_{COH}$≈200 bars (Table 1 and Figure 1), which begins to predominate over silicate vapor pressure ($P_{SV}$) below $T_{COH}$≈4,800 K. Particularly in the case of Earth, modeled temperatures are well beyond laboratory conditions where the thermochemical behavior of the system can be accessed experimentally. The associated uncertainty in silicate vapor pressure at such temperatures translates into uncertainty in the onset temperature for Earth's C-O-H atmosphere. Accordingly, we calculate models of terrestrial C-O-H atmospheres at a range of base temperatures $T_{COH}$≈4,750-5000 K to capture the behavior of this system despite this uncertainty. In summary, we calculate the structure of C-O-H-dominated atmospheres at a range of temperatures near 3,000 K (proto-lunar disk) and 5,000 K (post-giant impact Earth).

2.3. Atmospheric composition via equilibration with magma ocean liquids

The chemical composition of volatile element atmospheres are calculated using magma-vapor dissolution equilibria of C-O-H species with underlying magma oceans. The model includes vapor molecules and associated dissociation products ($CO_2$, CO, $H_2O$, $H_2$, $O_2$, OH, O and H). Hybrid silicate-volatile species (e.g., MgOH) are neglected because they are expected to be minor species at the relevant thermal states in the proto-lunar disk (Pahlevan et al., 2016). Because post-giant impact magma oceans are assumed to have BSE composition and to be metal-saturated, the oxygen fugacity ($fO_2$) of equilibration with the atmosphere is taken as 1-2 orders of magnitude below the iron-wüstite buffer ($\log fO_2$ = IW-2 to IW-1) (Frost, 1991; Frost et al., 2008; Hirschmann, 2012). For hydrogen and carbon, the dominant dissolved species in the silicate magma are expected to be $OH^-$ and $CO_3^{2-}$, and solubility laws relating these magmatic species to the partial pressure of $H_2O$ and $CO_2$ in the atmosphere are used (Pan et al., 1991; Sossi et al., 2023):

$$P_{H2O}(bars) = (2.9 \times 10^4)(x_L^H/0.01)^2 \tag{1}$$

$$P_{CO2}(bars) = (6.7 \times 10^6)(x_L^C) \tag{2}$$

where $x_L$ represents the mass fraction of the volatile element (H or C) in the magma. Both these solubility studies found a negligible dependence of solubility on temperature, supporting the application of these experimental results to the more extreme temperatures prevailing after a giant impact. Moreover, although the adopted $CO_2$ solubility law was experimentally derived for a basaltic liquid, $CO_2$ is expected to become more soluble in depolymerised ultramafic melts. Using the compositional dependence derived in (Shishkina et al., 2014), roughly proportional to NBO/T, the change in melt composition is estimated to increase $CO_2$ solubility by a factor 2-3 compared to the parameterization of (Pan et al., 1991). Given the stark contrast in the solubility behavior of H and C in the post giant-impact Earth and disk (see Table 1), the conclusions of this study are expected to be robust to such modest refinements in magma solubility of volatiles. With $P_{H2O}$ and $P_{CO2}$ constrained via dissolution equilibria, partial pressure of these species' reducing counterparts ($H_2$, CO) are calculated using gas-phase equilibria at the buffered oxygen fugacity and thermal dissociation is calculated using dissociation equilibria using standard thermodynamic data (Chase, 1998). The partial pressure of volatile inventories are related to atmospheric surface densities by incorporating the vertical redistribution effect for well-mixed multi-component atmospheres (Bower et al., 2019). At the thermal states of interest, magma oceans are expected to be fully molten and assumed fully convective. The calculations involve liquid-vapor equilibria and both the atmosphere and magma ocean are considered to be well-mixed, in chemical equilibrium, and internally chemically homogenous (see Appendix B for details).

2.4. Vertical atmospheric thermal and chemical structure

The vertical structure of volatile-rich atmospheres are calculated using reactive multi-species adiabats. Adiabatic structures are important in discussions of hydrodynamic outflows because they invoke no energy source for powering the escape other than the thermodynamic state – specifically the enthalpy – of the gas at the base of the atmosphere (Lamers and Cassinelli, 1999). Adiabatic structures are calculated by evaluating the entropy of the gas mixture at the atmospheric base in equilibrium with the magma assuming that the gas is an ideal mixture of ideal gases, an adequate approximation for a high-density, high-temperature gas (the terrestrial atmosphere, Fig. 1) and an excellent approximation for a low-density, high-temperature gas (the proto-lunar disk atmosphere, Fig. 2):

$$S = \sum_i x_i S_i(T) - R \sum_i x_i ln x_i - R ln P \quad (3)$$

where $x_i$ and $S_i$ are the mole fraction and specific molar entropy of gas species i, R is the ideal gas constant, and P is the pressure in bars. Once the entropy of the gas mixture at the basal thermal state is calculated, adiabatic structure is solved by recalculating gas-phase equilibrium at a lower pressure while keeping entropy and atomic composition of the parcel constant. This method of calculating gas speciation as a function of pressure was previously used to describe the structure of low-mass stars (Hayashi and Nakano, 1963). In this way, the chemical ($x_i$) and thermal (T) structure of adiabatic C-O-H atmospheres as a function of pressure (P) can be described. Results from these calculations are shown in Figures 1 and 2. To extract the value of the adiabatic index ($\Gamma$) from these calculations, we plot log P versus log ρ and use a linear regression to find the value of $\Gamma$ that gives the best-fit slope. The adiabatic index fully characterizes vertical temperature and chemical structure of model atmospheres, and is an important component for determining the onset of escape (§2.5) and for calculating outflow properties (§2.6, Table 2).

## 2.5. Assessing hydrodynamic stability of volatile element atmospheres

The existence and properties of a thermally-driven hydrodynamic outflow can be assessed once the composition and thermal structure of a given atmosphere is specified. Following the pioneering work on the solar wind (Parker, 1963), a widely used approach to characterize thermal structure in outflows is to adopt the polytropic equation of state ($P = K\rho^{\Gamma}$), with P the gas pressure, ρ the gas density, $\Gamma$ the polytropic index, and K a constant. The polytropic equation can describe isothermal ($\Gamma = \Gamma_{iso} = 1$) as well as adiabatic ($\Gamma = \Gamma_{ad} > 1$) thermal structures, and is thus suitable for assessing the hydrodynamic state of adiabatic atmospheres. For inert and non-reactive atmospheres in which the composition is everywhere fixed (such as the solar wind), the adiabatic value of the polytropic index $\Gamma_{ad}$ is equal to $C_p/C_v$ with $C_p$ and $C_v$ the specific heats at constant pressure and volume, respectively. For reactive atmospheres, such as the atmospheres we investigate here, the heat released by chemical reactions and changes in the number of particles due to such reactions must be incorporated into $\Gamma_{ad}$ (Hansen et al., 2012). The calculation procedure described above (§2.4) captures these effects, which turn out to be crucial to the conclusions. In an atmosphere whose structure is characterized by a single adiabatic exponent ($\Gamma=\Gamma_{ad}$), the condition for the onset of outflows can be expressed with a generalized escape parameter (Λ) (Parker, 1963):

$$\Lambda = \left(\frac{\Gamma-1}{\Gamma}\right)\left(\frac{GM\bar{\mu}_0}{rkT_0}\right) < \Lambda_{crit} \quad (4)$$

with G the gravitational constant, M the mass of the central body, $\bar{\mu}_0$ the mean molecular weight of the gas at the atmospheric base, r the distance to planetary center, k the Boltzmann constant, and $T_0$ the temperature at the atmospheric base. Because the mass of the proto-lunar disk is thought to be a few percent of the Earth's mass (Canup, 2004), to good approximation we can neglect disk self-gravity and exclusively focus on the effects of Earth's gravity when considering outflows from the circumplanetary disk. The generalized escape parameter ($\Lambda$) quantifies the capacity of an atmosphere to maintain hydrostatic equilibrium in the presence of pressure gradients and to remain gravitationally bound. Equation 4 shows that any atmosphere experiences outflow if sufficiently hot or sufficiently close to isothermal. An earlier study on hydrodynamic escape in the nascent Earth-Moon used the same criterion for escape ($\Lambda_{crit}$=1) for the post-giant impact Earth and proto-lunar disk (Genda and Abe, 2003). Here, motivated by more recent studies of hydrodynamic disk winds (Waters and Proga, 2012), we adopt a critical value for outflows launched from a disk ($\Lambda_{crit}$=2) that is twice that for outflows launched from a non-rotating planet ($\Lambda_{crit}$=1), incorporating Keplerian orbital motion, which prepossesses half the kinetic energy needed to unbind the gas and requires only half to be contributed thermally. This factor of two was previously invoked in an insightful and underappreciated work (Desch and Taylor, 2011) on which the current work builds.

2.6. Hydrodynamic structure of adiabatic disk atmospheres and associated mass loss rates

A model that yields the basal atmospheric temperature and basal mean molecular weight ($\lambda_0$), and the vertical variations of these parameters ($\Gamma$) is sufficient to determine the mechanical status (hydrostatic versus hydrodynamic) of an atmosphere in a given gravity field (see Fig. 3). For disk atmospheres in a state of hydrodynamic outflow, mass loss occurs at a finite rate, which can be calculated with these parameters ($\lambda_0$, $\Gamma$) using hydrodynamic descriptions analogous to those of the solar wind (Parker, 1963). To adapt a non-rotating spherically-symmetric Parker-wind to rotating outflows launched from an orbiting proto-lunar disk, we implement two modifications. First, we write the momentum equation as:

$$v\frac{dv}{dr}+\frac{1}{\rho}\frac{dP}{dr}+\frac{GM_p}{r^2}-\frac{j^2}{r^3}=0 \tag{5}$$

where, in addition to the gravitational acceleration induced by central planetary gravity, we include a centrifugal term in the momentum equation with $j^2 = GMr_0$ the angular momentum of the gas at the wind base $r_0$, which represents the centrifugal boost experienced by orbiting gaseous matter, a feature previously used in models of the photoevaporation of circumstellar disks (Adams et al., 2004). Second, in contrast to spherical winds where one-dimensional analysis suffices, a disk wind is an intrinsically two-dimensional problem, with the radial and vertical cylindrical components of the outflow not *a priori* determined (Waters and Proga, 2012). To make the problem tractable analytically, assumptions about flow geometry are needed. A simple approximation for marginally unbound disk winds (basal Mach number much less than unity) is that the outflow is dominated by radial expansion from the disk edge where the gravity is weakest, the so-called subcritical regime (Adams et al., 2004). This regime may be particularly relevant to the proposed outflows because we invoke hydrogen-rich disk atmospheres that are encountered towards the end of proto-lunar disk condensation, and the proto-lunar disk is expected to cool and condense from large radii inwards, triggering a wave of outflows whose wind base propagates inwards. In this regime, the outflow geometry is radial, subtending a constant solid angle during outward expansion, and the continuity equation has the form:

$$\dot{M} = 4\pi r^2 \rho(r) v(r) F \tag{6}$$

with ρ the gas density, v the outflow velocity, and F a filling factor (<1), assumed to be constant. With this approximation, equations governing the outflow reduce to one-dimensional form as the continuity equation together with the momentum equation are sufficient to calculate hydrodynamic solutions and associated mass loss rates. The procedure for using Equations 5 and 6 together with atmospheric input parameters ($\lambda_0$, $\Gamma$) to derive hydrodynamic disk wind solutions is described in the Appendix E. Using this procedure, the outflow properties of disk atmospheres are calculated and shown in Figure 4 and Table 2.

## 3. Results

### 3.1. Composition and structure of the volatile element terrestrial atmosphere

The volatile-rich terrestrial atmosphere in equilibrium with an underlying metal-saturated magma ocean of bulk silicate Earth composition is composed primarily of CO with secondary H-$H_2$ whereas $CO_2$ and $H_2O$ are minor species throughout the atmosphere (Figure 1a). Importantly, because H is highly soluble as $H_2O$, most (>96%) exchangeable hydrogen (magma ocean plus atmosphere) remains dissolved in the magma whereas most (>99%) carbon is outgassed into the earliest terrestrial atmosphere as CO (Table 1). The predominance of carbon monoxide is important for the structure and hydrodynamic stability of this atmosphere. First, the high mean molecular weight of the atmosphere – and thus small scale height relative to Earth's radius ($H_E \approx 150$-$200$ km, Table 1) – is heavily influenced by the prevalence of CO, producing a compact reservoir strongly gravitationally bound to Earth. Second, the thermodynamic stability of CO ($D_{CO}=1080$ kJ/mol) – which experiences negligible thermal dissociation and is nearly inert at all altitudes – results in a cool upper atmosphere relative to the temperature at the base (Figure 1b). Vertical temperature gradients are characterized with an adiabatic index ($\Gamma_{ad}>1$) and its departure from the isothermal value ($\Gamma_{iso}=1$) (§2.4). The adiabatic index of an inert CO-dominated atmosphere ($\Gamma_{ad}=1.27$) produces cool temperatures and smaller scale heights at altitude, an effect that further stabilizes the atmosphere against expansion and outflow. The combination of a high mean molecular weight exsolved gas, steep vertical thermal gradients, and a deep position in the gravitational potential well of the Earth places the post-silicate terrestrial atmosphere firmly in the regime of a hydrostatic and gravitationally-bound atmosphere throughout its cooling history (see §3.3 and Figure 3a).

3.2. Composition and structure of the volatile element proto-lunar disk atmosphere

The volatile element-dominated proto-lunar disk atmosphere in equilibrium with an underlying metal-saturated magma layer is everywhere composed primarily of H-$H_2$ and subordinate CO whereas $H_2O$ and $CO_2$ only appear as minor species (Figure 2a). In contrast to the terrestrial case, the lower surface density and lower gravitational acceleration in the proto-lunar disk (≈0.02x Earth gravity, see Table 1) results in nearly complete outgassing of carbon *and* hydrogen from the disk magma into the enveloping atmosphere (>97% H outgassed in the case plotted in Figure 2, see Table 1). As a result, the H/C atomic ratio of the disk atmosphere primarily reflects the bulk H/C of the disk as a whole, i.e., H/C=12 counting atoms for a BSE-composition disk (Hirschmann and Dasgupta, 2009). The hydrogen-rich nature of the exsolved gas has consequences for the structure and hydrodynamical stability of the post-silicate proto-lunar disk atmosphere and therefore the lunar composition. First, the low mean molecular weight of the hydrogen-rich C-O-H gas produces an inflated disk scale height ($H_D=[2k_BTr^3/GM\mu]^{1/2}\approx 8{,}000$ km at the Roche radius at 3 $R_E$, Table 1), a property shared with atmospheres undergoing outflow. Second, because reactive atomic hydrogen (H) is a major species in the lower disk atmosphere (Figure 2a), adiabatic ascent is accompanied by equilibrium $H_2$ recombination, a reaction that releases a large quantity of heat into the upper disk atmosphere (434 kJ/mol). Release of such quantities of heat in ascending parcels thwarts adiabatic cooling, making the H-$H_2$-dominated adiabat nearly isothermal vertically (Figure 2b), a property that inflates the upper atmosphere and is known to activate outflows (Parker, 1963). The combination of a low mean molecular weight atmosphere, a nearly isothermal vertical structure, and a relatively shallow position in the Earth's gravitational potential well places the disk atmosphere beyond the threshold of hydrostatic breakdown and into the regime of thermally-driven hydrodynamic outflow, even in the absence of external high-energy solar or stellar photons (see §3.3 and Figure 3b).

### 3.3. Hydrodynamic stability of volatile element atmospheres of Earth and the proto-lunar disk

We assess hydrodynamic stability with respect to thermally-driven expansion ("blowoff") of C-O-H atmospheres in contact with metal-saturated magma oceans. We find that such atmospheres are hydrostatically bound on the post-giant impact Earth and in a state of hydrodynamic outflow from the proto-lunar disk. The generalized escape parameter ($\Lambda$) (see Eqn. 4) as a function of basal atmospheric temperature – which represents different epochs in atmospheric cooling history – is evaluated for the two atmospheres and illustrated in Figure 3. Generalized escape parameter values greater than the critical value ($\Lambda > \Lambda_{crit}$) indicate hydrostatic atmospheres in which the atmospheric pressure goes to zero at finite altitude, whereas lower than critical generalized escape parameter values ($\Lambda < \Lambda_{crit}$) describe atmospheres with finite pressure at infinity, that is, atmospheres that cannot maintain hydrostatic equilibrium in a vacuum and must thermally expand, producing a planetary wind analogous to the solar wind. For the case of the Earth, a high molecular weight carbon-rich gas composition, a relatively cool upper atmosphere, and an intrinsically strong gravity field combine to produce a compact atmosphere far from the threshold of hydrodynamic outflow (Figure 3a). Closed-system behavior of the Earth's volatiles in the aftermath of the Moon-forming giant impact is implied and is consistent with the empirical stable isotopic record (Fischer et al., 2024; Greenwood et al., 2018). Despite some uncertainty in the onset temperature of the terrestrial C-O-H atmosphere (§2.2), the observed temperature-dependence of the mechanical status of the terrestrial atmosphere (Figure 3a) together with the requirement that onset temperature be lower than the critical temperature for BSE-composition material estimated to be ~6,000 K (Caracas, 2024) suggests robustness of these results to improvements in thermochemical data. In the proto-lunar disk, by contrast, a low molecular weight hydrogen-rich gas composition, an upper atmosphere prevented from strong cooling via $H_2$ recombination, and a shallow position in Earth's potential well combine to produce an atmosphere sufficiently inflated to overcome the confining gravity of Earth and to hydrodynamically shed mass to interplanetary space. In contrast to recent assessments (Nakajima and Stevenson, 2018; Pahlevan et al., 2016), a hydrodynamic outflow from a Roche-interior ($< 3\ R_E$) proto-lunar disk is indicated, on the basis of the expectation that $H_2$ – not $H_2O$ as previously supposed – was the primary species in the disk atmosphere (Figure 3b). These results suggest that the nascent Earth-Moon system developed a cometary tail composed of proto-lunar volatiles during the lifetime of the melt-vapor disk and that the lunar complement of volatile elements need not have been redistributed to the Earth but may have been expelled to interplanetary space. Indeed, the substantial gap between the threshold of hydrostatic breakdown and the conditions prevailing in a pure C-O-H atmosphere (Figure 3b) suggests that outflows would have set in before complete condensation of silicate species and could have entrained a substantial inventory of moderately volatile elements in the outflow via gas collisions. These results demonstrate a dichotomy in the hydrodynamic behavior and fate of volatile element atmospheres on Earth (fully retained) and proto-lunar disk (partially lost), consistent with their respective chemical and isotopic records.

3.4. Lunar sodium abundance and the radial extent of the proto-lunar disk

By mass, the most abundant moderately volatile element (MVE) in lunar rocks is sodium, which constitutes ≈0.3 wt% of the bulk silicate Earth (McDonough and Sun, 1995) but only ≈0.03 wt% of the bulk silicate Moon (Ringwood and Kesson, 1977). Accordingly, the vaporized Na abundance is expected to exert strong influence on the structure and dynamics of volatile element-dominated disk atmospheres. To assess the influence of MVEs on the proposed outflows, we evaluate the properties ($\lambda_0$, $\Gamma$) of several adiabatic C-O-H-Na atmospheres at two disk radii, with the fraction of sodium vaporized and entrained taken as a free parameter (Table 2). Vaporized sodium influences the atmosphere in two ways. First, Na increases the mean molecular weight of the C-O-H gas mixture and thus greater sodium translates to higher gravitational binding ($\lambda_0$). Second, because Na vapor is expected to be mainly monatomic throughout the atmosphere, it is effectively inert, increasing the adiabatic index ($\Gamma$=1.125−1.20, Table 2) above the value for pure C-O-H atmospheres ($\Gamma$=1.11, Figure 2). Both effects cause the upper atmosphere to contract and the vigor of outflows to be inversely dependent on the vaporized Na abundance. To obtain some quantitative sense of how Na encumbers outflows – or conversely how much Na can be entrained – we evaluate the hydrodynamic status of C-O-H-Na atmospheres using Equation 4 and resolve the transition from hydrostatic to hydrodynamic. The precise amount of Na that can be vaporized without halting the outflow depends on the depth of the gravity well, with 25-60% of the total BSE-scaled abundance entrained for a disk wind launched from 2-3 Earth radii (Figure 4). These distances represent locations in a circumterrestrial disk from which proto-lunar material is sourced in disk models of lunar accretion (Canup et al., 2015). This model therefore suggests a scenario in which magmas at greater planetocentric distances are more volatile depleted whereas more central magmas are more volatile-rich. Although a complete assessment of lunar Na loss would require evolutionary calculations ranging from BSE to BSM compositions, these onset calculations demonstrate that substantial Na loss from the proto-lunar disk is plausible. The mass loss rate in an adiabatic outflow can be calculated with an assumption about the geometry of the outflow and can be characterized by the Mach number at the base (see §2.6 and Appendix E). The timescale associated with such outflows ($\tau \sim 3R_E$/basal outflow velocity) is of order one *week* (Table 2), much shorter than the secular cooling timescale of the disk measured in years (Pahlevan and Stevenson, 2007; Thompson and Stevenson, 1988). This rapidity suggests that some other process – such as H exhaustion – may ultimately determine the fraction of sodium and other MVEs retained or lost. Relatedly, partial condensation of sodium and other MVEs may occur along the cooling flow. The influence of suspended droplets or dust can be incorporated by including condensate mass in calculations of the mean molecular weight of the fluid (Youdin and Goodman, 2005). In this way, the influence of MVEs – whether vaporized or entrained as droplets or dust – is captured by the calculations shown in Figure 4. Finally, recent works quantifying isotopic fractionation of lunar volatiles (e.g., K, Rb, Zn) infer that evaporation into a ~99% saturated vapor medium is required to reproduce the data (Dauphas et al., 2022; Nie and Dauphas, 2019). Such a condition naturally arises for an outflow with basal Mach number ~0.01 (Pritchard and Stevenson, 2000) and is encompassed in our solutions (Figure 4). Such a kinetically-controlled evaporation process does not preclude other isotopic signatures, e.g., due to the loss of equilibrium vapor (Wang et al., 2019), from being imprinted onto proto-lunar liquids.

## 4. Discussion

### 4.1. Implications for the origin and abundance of endogenous lunar water

Following the initial analysis of Apollo samples, the lunar interior was considered "bone dry" with $H_2O$ contents < 1 ppb by weight. In a pioneering study, magmatic water and other volatiles from the lunar interior were discovered in volcanic glasses formed in pyroclastic eruptions (Saal et al., 2008), prompting ongoing reevaluation of the bulk lunar H abundances now measured in ppm (McCubbin et al., 2023). On the theoretical side, $H_2O$-rich proto-lunar disk atmospheres were found to be gravitationally bound by the Earth (Nakajima and Stevenson, 2018) and equilibrium dissolution of water vapor into proto-lunar disk liquids was found to introduce only 1-2 orders of magnitude of depletion relative to the bulk abundances in the disk due to the high solubility of water in silicate magma (Pahlevan et al., 2016). Here, we have considered a more realistic $H_2$-rich composition for the proto-lunar disk atmosphere and found that such an atmosphere is unbound and likely to be in a state of escape, implying that the 1-2 orders of magnitude water depletion is a lower limit to the primordial lunar water depletion. In the context of our model, the lunar water and sodium depletions are coupled because both result from hydrodynamic blowoff of a volatile-rich disk atmosphere. Although partial Na retention (§3.4) implies partial H retention, massive Na loss – 80-90% depletion relative to the BSE (Ringwood and Kesson, 1977) – implies at least as extensive hydrogen loss from disk atmosphere and coexisting magmas, limiting the expected magnitude of water dissolved into proto-lunar magmas. As an example, 100 ppm H in the proto-lunar disk produces 2.3-4.3 ppm H dissolved in proto-lunar liquids (Table 1), corresponding to 21-39 ppm $H_2O$ even before hydrodynamic escape. Making the conservative assumption that lunar H depletion is only as extensive as Na depletion (~1 order of magnitude) yields 2-4 ppm $H_2O$ for the bulk Moon, which is in the lower end of the range of estimates (McCubbin et al., 2023). D/H fractionation accompanying such loss is likely negligible on the magnitude of interest (see Appendix F). Finally, dissolved H/C in proto-lunar magmas is predicted to have been elevated by many orders of magnitude over the BSE ratio (see Table 1) due to the solubility contrast for H and C (Hirschmann, 2012). By contrast, the inferred H/C of the lunar volcanic glass magmas is similar to undegassed MORB (Wetzel et al., 2015), arguing against disk dissolution as the primary volatile source for lunar volcanic glasses. A viable alternative is efficient hydrodynamic removal of disk volatiles followed by later resupply by asteroidal and/or cometary impacts (Cassata et al., 2025). The retention of terrestrial volatiles (§3.3 and Figure 3) may be why the signature of late accretion is not as discernible on Earth against a more volatile-rich background. In summary, equilibrium disk dissolution can produce ppm levels of water in the bulk Moon with unfractionated D/H and accompanied by essentially no carbon, and late accretion of lunar water may be required.

## 5. Conclusions

The Moon-forming giant impact subjected the terrestrial and lunar material to temperatures high enough to partially vaporize the major constituents. Following condensation of the major silicate- and metal-forming elements (Mg, Si, Fe), two distinct volatile-element enriched atmospheres arose on post-impact Earth and the proto-lunar disk. Such atmospheres are promising environments for a highly selective removal of volatile elements, including the lunar volatile depletion first inferred from analysis of Apollo samples. Novel conclusions from this study include: (1) Massive water dissolution in a deep magma ocean resulted in a CO-dominated post-giant impact terrestrial atmosphere that was strongly bound gravitationally, leading to retention of terrestrial volatiles in the face of the thermal extremes of the giant impact, a crucial factor in the subsequent habitability of Earth; (2) On the proto-lunar disk, large-scale exsolution of both H and C produced a C-O-H atmosphere dominated by atomic hydrogen (H) at depth and molecular hydrogen ($H_2$) at altitude. This composition and thermal state resulted in an atmosphere sufficiently inflated to overcome the confining gravity of Earth and shed proto-lunar volatiles into interplanetary space; (3) Expected outflows from the proto-lunar disk are strong enough to propel volatiles – including the main mass contributor, sodium – from a Roche-interior ($r < 3R_E$) disk out to interplanetary space. Disk atmospheres at larger radial distances shed more volatiles whereas atmospheres in the inner disk retain relatively more; (4) Lunar volatile abundances (e.g., Na) and isotopic compositions (e.g., K, Rb, Zn) are diagnostic of the radial structure of the proto-lunar disk towards the end of silicate condensation. Such a link between models and observations may yield new constraints on viable proto-lunar disk models that can reproduce the lunar composition.

**Appendix A – Onset temperatures of volatile element (C-O-H-Na-dominated) atmospheres**

Models of the proto-lunar disk typically involve a circumplanetary disk composed of silicate melt and co-existing silicate vapor (Canup et al., 2015; Nakajima and Stevenson, 2014; Nakajima and Stevenson, 2018; Pahlevan et al., 2016; Thompson and Stevenson, 1988). However, as the proto-lunar disk cools and the silicate vapor condenses, the residual atmosphere becomes enriched in volatile elements. Such volatile element-rich atmospheres are successors to the silicate vapor atmospheres that prevail during earlier stages of the disk evolution. An important question is the hottest temperature at which such volatile-dominated atmospheres appear, because thermally-driven outflows require an atmosphere that is sufficiently hot and extended for the outer layers to overcome confining gravity. To estimate the onset temperature of C-O-H-dominated atmospheres, we adopt the criterion:

$$P_{COH} = P_{SV} \tag{A1}$$

with $P_{COH}$ and $P_{SV}$ the partial pressure of volatile (C, O, H) and major element silicate (Mg, Si, Fe) bearing gaseous species, respectively. The rationale for this criterion is that when a condensible is present in an atmosphere below a certain molar abundance, adiabatic expansion condenses most of that component out during adiabatic ascent (Ingersoll, 1969) such that the upper atmosphere – from which outflows are sourced – can be modeled as a nearly pure volatile system. Because magma solubility of highly volatile elements H and C is nearly temperature-independent (Equations 1 and 2) but silicate vapor pressure is highly temperature-dependent, for a given C-O-H volatile inventory and pressure ($P_{COH}$), criterion A1 corresponds to a temperature below which the post-giant impact atmosphere is volatile-dominated ($T_{COH}$). Approximating the silicates using a previously described thermodynamic model with BSE major element composition (molar Fe/Mg ~ 0.1) (Pahlevan et al., 2016), silicate vapor pressure can be expressed as a function of temperature and fit to an equation of the form $P_{SV}=P_0\exp(-T_0/T)$ with $P_0=4.1\text{x}10^7$ bars and $T_0=5.8\text{x}10^4$ K. Similarly, if the magma ocean is metal-saturated (e.g., due to a suspension of metallic droplets kept aloft via convection), the co-existing vapor would also be saturated with respect to Fe-rich metallic vapor. Metallic Fe vapor pressure can be expressed with an equation of the same form with $P_0=2.3\text{x}10^6$ bars and $T_0=4.6\text{x}10^4$ K. Although the Clausius-Clapyron relation breaks down at temperatures near the critical point, this expression yields a first-order estimate of silicate (and metallic) vapor pressure as a function of temperature far from the critical point, estimated to occur at $T_C=6{,}000$ K for BSE-composition material (Caracas, 2024). For $P_{COH} \approx 0.042$ bars appropriate for the proto-lunar disk (see Table 1), criterion O1 yields $T_{COH} \approx 2800$ K, demonstrating the appropriateness of this thermal state for describing C-O-H atmospheres experienced by the disk. For the post-giant impact Earth, $P_{COH} \approx 200$ bars (Table 1) surpasses silicate vapor pressure below $T_{COH} \approx 4740$ K, justifying this thermal state in describing Earth's hottest C-O-H atmosphere. Due to uncertainties in the thermochemical behavior of BSE-composition material at these extreme temperatures, we consider a range ($T_{COH}$≈4,750-5,000 K) for the onset temperatures of Earth's C-O-H atmospheres to constrain the temperature-dependence of the Earth's behavior in this epoch.

A number of refinements to this simple estimate can be considered for the disk. First, because moderately volatile elements (dominated by Na) participated in the lunar volatile loss episode, their partial vapor pressure could be included on the left-handed side of A1. Counting atoms, Na loss was comparable to H loss if the proto-lunar disk started with BSE composition and ended up with BSM composition (Assuming BSE composition, the number of proto-lunar Na atoms – most of them lost – is proportional to 2700 ppm/23 amu comparable to 100 ppm/1 amu which is proportional to the number of H atoms inherited and lost). Inclusion of Na can therefore increase $P_{COH}$ by up to a factor of two. Second, throughout this work, we use a surface density derived from uniformly spreading two lunar masses between 1-5 Earth radii (=$5x10^7$ kg/m$^2$). Since the pressure exerted by the C-O-H atmosphere is proportional to the surface density of this atmosphere, more massive and more compact disks would exhibit greater pressure from outgassed volatile species. Variations in surface density and thus $P_{COH}$ of a factor ≈4 (derived by uniformly spreading 3 lunar masses from 1-3 Earth radii relative to our reference case) are considered within the plausible range. Third, we have taken 3 Earth radii as the reference standard for all disk calculations, but the proto-lunar disk would have exhibited radial extent, such that the effective gravity experienced by disk material (Table 1) – and thus the pressure exerted by the atmospheric overburden – would have similarly exhibited a range, increasing about 80% at 2 Earth radii relative to the reference at 3 Earth radii. Considering these factors together, the pressure exerted by the volatile element atmosphere ($P_{COH}$) could have been an order of magnitude or more higher than our reference case. The shaded area in Figure A1 illustrates the plausible range of volatile element pressures ($P_{COH}$=0.042-0.60 bars) expected for the proto-lunar disk on this basis, which translates to a range of onset temperatures ($T_{COH}$≈2,800-3,200 K) for volatile-dominated atmospheres, which we adopt for proto-lunar disk calculations (Table 1 and Figure 3).

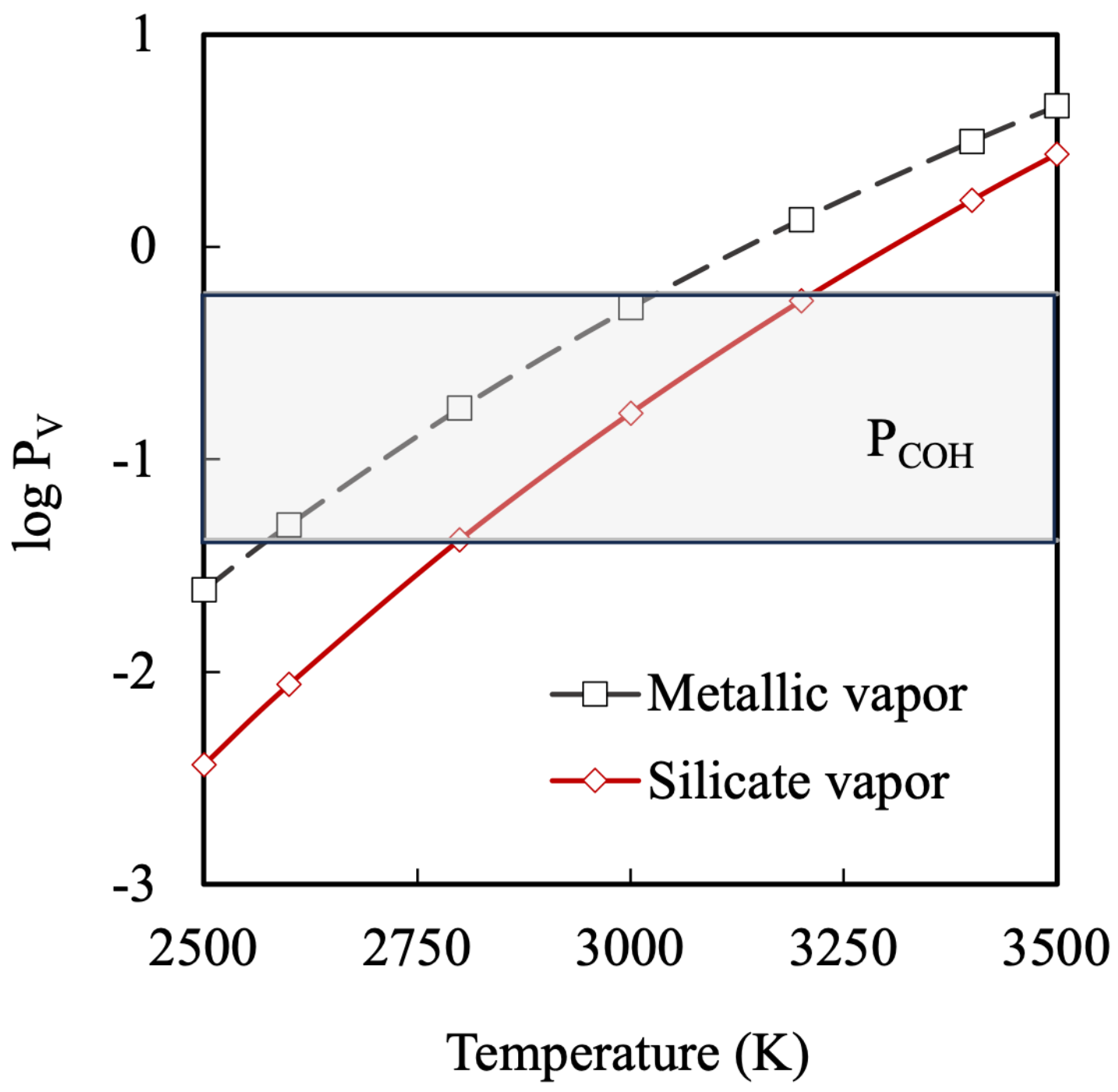

**Figure A1 – Onset temperatures of volatile element (C-O-H-Na) disk atmospheres.** The vapor pressure of silicate vapor (modeled using an olivine thermodynamics, Pahlevan et al. 2016) and metallic vapor (modeled with the vapor pressure of pure liquid Fe) versus temperature. The shaded region labeled $P_{COH}$ represents the range of pressures exerted by volatile elements for plausible models of a BSE-composition proto-lunar disk. $P_{COH}$ exceeds the silicate vapor pressure for T=2,800-3,200 K, justifying these thermal states for the application of a C-O-H volatile model. The metallic vapor pressure curve is dashed because it assumes that metallic Fe vapor is saturated but liquid metallic Fe is expected to condense at the midplane over this temperature range (see Appendix D), making the saturation vapor pressure curve an upper limit on the atomic Fe pressure.

## Appendix B. Magma ocean atmospheric compositions in the C-O-H system

The procedure for calculating the chemical composition of C-O-H atmospheres overlying metal-saturated magma oceans relies on equilibrium thermodynamics and hydrostatic balance, which is a good approximation even in hydrodynamic atmospheres so long as the basal Mach number is much less than unity. We are interested in post-giant impact thermal states that are cool enough for metallic Fe to have condensed, such that the fugacity or partial pressure of $O_2$ at any temperature will be related to the iron-wüstite (IW) mineral buffer:

$$2FeO(s) = 2Fe(s) + O_2(g) \tag{B1}$$

$$f_{O2}^{IW}(T) = \exp(-\Delta S_1^0/R)\exp(-\Delta H_1^0/RT) \tag{B2}$$

with $\Delta S_1^0 = 0.127\ kJ/mol.K$ and $\Delta H_1^0 = 541.0\ kJ/mol$ from the JANAF tables (Chase, 1998) and the universal gas constant $R = 0.008314\ kJ/mol$.K. The precise value of $P_{O2}$ will depend on the precise composition of the silicate and metallic phases, for example, the FeO content of the melt. We consider $P_{O2}$ 1-2 log units below the IW buffer, consistent with the redox state of the lunar interior as derived from sample studies (Wadhwa, 2008). In our model, the oxygen buffering capacity of the liquids dominates that of the vapors and the $fO_2$ of the magma ocean is assumed to be imposed onto the atmosphere, an approach that has previously been adopted for magma ocean atmospheres of the terrestrial planets (Hirschmann, 2012). An alternative assumption is that the buffering capacity of the gas overwhelms that of liquids, in which case the atmosphere imposes its $fO_2$ onto the magma ocean, an approach that may be valid for planetary atmospheres that capture substantial hydrogen inventories from surrounding nebulae (Charnoz et al., 2023). Knowing the fugacity of oxygen is important because this element reacts with all other species and influences the overall gas phase equilibrium speciation. In addition to $O_2$, whose partial pressure is determined by equilibrium between liquids, we consider a number of volatile molecular and atomic species ($CO_2$, CO, $H_2O$, $H_2$, OH, O and H) expected to be present in post-giant impact C-O-H atmospheres. Due to dissolution equilibria between the magma ocean and atmosphere, the partial pressure of $CO_2$ and $H_2O$ can be related to liquid concentrations via solubility laws given in the main text (§2.3). Partial pressures of CO, $H_2$, H are related to those of $CO_2$, $H_2O$, and OH via redox reactions:

$$CO_2(g) = CO(g) + \frac{1}{2}O_2(g) \tag{B3}$$

$$K_3 = \frac{{p_{O2}}^{1/2} p_{CO}}{p_{CO2}} = \exp(-\Delta S_3^0/R)\exp(-\Delta H_3^0/RT) \tag{B4}$$

whereas the partial pressures of H and O are related to that of $H_2$ and $O_2$ via dissociation equilibria. $CO_2$ (dissociation energy $D_{CO2}$=530 kJ/mol), $O_2$ ($D_{O2}$=502 kJ/mol), $H_2O$ ($D_{H2O}$=502 kJ/mol), $H_2$ ($D_{H2}$=434 kJ/mol), and OH ($D_{OH}$=425 kJ/mol) all experience significant thermal dissociation whereas CO ($D_{CO}$=1080 kJ/mol) experiences negligible thermal dissociation in this environment. Thermodynamic data used to relate equilibrium partial pressures are adopted from JANAF tables (Chase, 1998) and given in Table S1. The adopted thermodynamic parameters are derived by fitting JANAF data from 0-3,000 K with a constant ΔH° and ΔS° and applying them to the proto-lunar disk (T≈3,000 K) and the post-giant impact Earth (T≈5,000 K). Application to the post-giant impact Earth requires a considerable extrapolation and should be considered more exploratory. One measure of the robustness of the results is the comparison to analogous calculations done with Atmodeller (Bower et al., 2025) (see Table B3). Despite the high partial pressure of CO in Earth's atmosphere (Table 1), graphite activity ($a_C$) is ~$10^{-3}$ ($a_C$=1 for graphite saturation) demonstrating that CO is stable with respect to graphite formation via CO(g)→C(s)+$O_2$(g). The same stability with respect to graphite applies *a fortiori* in the proto-lunar disk. In addition to liquid-liquid-vapor, liquid-vapor and vapor-vapor equilibria, we impose constraints derived from the purity of the system by assuming that no species are present other than those that are accounted for explicitly:

$$\sum_i x_i = 1 \quad \text{(B5)}$$

and we characterize the thermal state of the equilibrium system using basal atmospheric entropy:

$$S = \sum_i x_i S_i(T) - R\sum_i x_i ln x_i - RlnP \quad \text{(B6)}$$

where $x_i$ is the mole fraction or volume mixing ratio, $S_i$ is the specific entropy of gas species i as a function of temperature, R is the universal gas constant, and P is the total pressure. The data for the entropy of the pure gases is derived from the JANAF tables (Chase, 1998) and shown in Table S2. We seek solutions in which the basal temperature corresponds to the temperature at which the atmosphere becomes volatile-dominated (§2.2). Basal atmosphere entropy is a useful measure of the thermal state because it can also be used to calculate adiabats and to characterize vertical atmospheric structure (§2.4). The vertical scale height and the mechanical status of magma ocean atmospheres are influenced by the mean molecular weight (§2.5), which we therefore track:

$$\bar{\mu} = \sum_i x_i \, \mu_i \quad \text{(B7)}$$

For well-mixed multi-component atmospheres, the relationship between the partial pressure and the surface densities of volatile inventories are modulated by the ratio of molecular weight of the individual gas species and the gas mixture as a whole (Bower et al., 2019):

$$P_i = \sum_i \sigma_i g_{eff} \, \bar{\mu}/\mu_i \quad \text{(B8)}$$

An equation like E8 is written for each of six H- or C-bearing species considered ($CO_2$, CO, $H_2O$, $H_2$, OH, and H) but $O_2$ and O are excluded. Unlike H and C, oxygen-bearing gas species are subject to an effectively infinite buffering capacity by coexisting liquids (see E1, E2) and therefore have atmospheric inventories that can be determined without explicit calculation of liquid-vapor partitioning. To link the surface densities of individual gaseous species to the total gaseous surface densities of H and C, we sum over H- and C-bearing species:

$$\sigma_V^{H\ or\ C} = \sum_i y_i \sigma_i \quad \text{(B9)}$$

where $y_i$ is the mass fraction of the species composed of the volatile element in question (e.g., 1/17 for H in HO) and the sum is over $H_2O$, OH, $H_2$, and H for hydrogen and $CO_2$ and CO for carbon. To link these atmospheric volatile element surface densities to equivalent surface densities in the magma layer, we impose mass balance:

$$\sigma_T^{H\ or\ C} = \sigma_L x_L^{H\ or\ C} + \sigma_V^{H\ or\ C} \quad \text{(B10)}$$

where $\sigma_L$ is the surface density of the magma layer and is derived by uniformly spreading 2 lunar masses of material in a disk extending from 1-5 Earth radii ($\sigma_L = 5 \times 10^7 kg/m^2$). There may be variations in surface density with radius and time, variations we do not explicitly calculate. Finally, we specify the value of the gravitational acceleration used in E8:

$$g_{eff} = \begin{cases} 10\ m/s^2,\ post - impact\ Earth \\ a_0\Omega/\sqrt{2\pi},\ proto - lunar\ disk \end{cases} \quad \text{(B11)}$$

where $a_0$ is the isothermal sound speed at the midplane ($= \sqrt{kT_0/\mu_0}$) and $\Omega$ is the Keplerian angular velocity at the disk radius ($r_0$) under consideration ($= \sqrt{GM/r_0^3}$). The derivation of this expression for effective vertical gravity in a disk can be found in Equations 1-6 of (Pahlevan et al., 2016). For the post-giant impact Earth, rapid rotation could lower the effective gravity at the surface and change the partitioning of the volatile elements relative to that calculated here. We do not pursue such refinements here. For the proto-lunar disk, because the vertical component of central gravity increases with altitude, the value of the effective gravitational acceleration is not an input parameter but depends on the mean molecular weight of the atmosphere and therefore on the output of the partitioning calculation. Therefore, the value of effective gravity must be solved simultaneously with the partitioning calculation. The derived values of effective vertical gravity at the Roche radius are in the range of 0.01-0.1 of the gravitational acceleration at Earth's surface (Table 1). By contrast, the gravitational acceleration due to the self-gravity of the proto-lunar disk ($g_{PLD} = 2\pi G\sigma_L$ ) is ≈0.002 x the acceleration at Earth's surface, justifying the neglect of disk self-gravity and a treatment exclusively focused on the gravitational field of the Earth. Using Equations 1 and 2 (from §2.3), E2, E4 (and 4 others corresponding to gas-phase reactions given in Table B1), E5, E6, E7, E8 (and five others analogous to E8), E9 and E10 (once for H and once for C), and E11, and the input parameters for surface density and radius of the disk under consideration ($\sigma_L, \sigma_T^H, \sigma_T^C, S, r_0$), we simultaneously solve for the temperature (T) and pressure (P) of liquid-vapor equilibration, as well as the partial pressures of eight C-O-H gaseous species, atmospheric surface density for six H-and-C-bearing gases, the total surface density of atmospheric H and C, the base atmospheric mean molecular weight, the effective vertical component of gravity (for disk calculations), and the mass fraction of dissolved hydrogen ($x_L^H$) carbon C ($x_L^C$) in the co-existing magma. Outputs from these calculations are shown in Table 1 and Table S3 and constitute basal conditions in Figures 1 and 2.

**Table B1** – The standard entropy and enthalpy of reactions (ln K = - ΔG°/RT, ΔG° = ΔH° - TΔS°) that determine the C-O-H speciation in high-temperature atmospheres. The adopted thermodynamic parameters are linear regressions to JANAF table data.

| Reaction | ΔS° (J/mol.K) | ΔH° (kJ/mol) |
|---|---|---|
| $CO_2$ (g) = CO (g) + ½ $O_2$ (g) | 85.3 | 281.5 |
| $H_2O$ (g) = $H_2$ (g) + ½ $O_2$ (g) | 56.1 | 247.2 |
| OH (g) = H (g) + ½ $O_2$ (g) | 43.8 | 184.2 |
| $H_2$ (g) = 2 H (g) | 116.1 | 444.4 |
| $O_2$ (g) = 2 O (g) | 131.2 | 504.7 |
| CO (g) = C (s) + ½ $O_2$ (g) | -85.8 | 113.4 |

**Table B2** – Entropy data for pure gases at 1 bar (S = A.lnT + B).

| Species | A (J/mol.K) | B (J/mol.K) |
|---|---|---|
| H (g) | 20.83 | -4.07 |
| O (g) | 21.0 | +41.9 |
| $H_2$ (g) | 31.4 | -49.9 |
| $O_2$ (g) | 35.3 | +0.766 |
| OH (g) | 32.1 | -1.31 |
| $H_2O$ (g) | 34.6 | -8.43 |
| $CO_2$ (g) | 54.7 | -105.9 |
| CO (g) | 33.75 | +2.5 |

**Table B3** – Calculated base composition of C-O-H atmospheres for the Earth and the proto-lunar disk. $T_0$ – atmospheric base temperature (K). ΔIW – base log$fO_2$ relative to iron-wüstite buffer. $P_{COH}$ – base COH pressure (bars). $P_i$ – partial pressure of gaseous species i at the base (bars).

| | $T_0$ (K) | ΔIW | $P_{COH}$ | $P_H$ | $P_{H2}$ | $P_{H2O}$ | $P_{CO}$ | $P_{CO2}$ |
|---|---|---|---|---|---|---|---|---|
| Earth | 4980 | -2 | 196 | 22.1 | 19.3 | 2.68 | 146 | 1.60 |
| | 5019 | -1 | 202 | 13.2 | 6.34 | 2.79 | 162 | 4.84 |
| | 4757 | -2 | 195 | 17.5 | 20.0 | 2.70 | 150 | 1.44 |
| | 4740 | -1 | 198 | 9.84 | 6.59 | 2.81 | 166 | 5.05 |
| Earth† | 5000 | -2 | 183 | 17.0 | 6.94 | 0.69 | 155 | 1.59 |
| | 5000 | -1 | 187 | 11.5 | 3.19 | 1.00 | 160 | 5.20 |
| Disk | 3000 | -2 | 0.0419 | 0.0187 | 0.0160 | 1.53E-3 | 4.58E-3 | 5.17E-5 |
| | 3000 | -1 | 0.0481 | 0.0185 | 0.0159 | 4.81E-3 | 5.19E-3 | 1.85E-4 |
| | 2800 | -2 | 0.0377 | 0.0112 | 0.0197 | 1.77E-3 | 4.59E-3 | 5.34E-5 |
| | 2800 | -1 | 0.0419 | 0.0108 | 0.0192 | 5.43E-3 | 5.12E-3 | 1.88E-4 |

The raw data for a subset of atmospheres hydrodynamically assessed in Figure 3.

In all cases the total system volatile mass fractions are [H] = 100 ppm, [C] = 100 ppm.

The adopted surface density for all the calculated disk models is $\sigma_T = 5 \times 10^7 kg/m^2$

Proto-lunar disk calculations conducted near the Roche radius at 3 Earth radii.

†Calculations done with Atmodeller (Bower et al. 2025) for an Earth-mass planet, demonstrating the robustness of the CO-dominated terrestrial atmosphere under identical modeling conditions.

## Appendix C. Calculation of the adiabatic index (Γ) and a sensitivity analysis

The vertical thermal and chemical structure of an atmosphere is a critical quantity in determining the hydrodynamic status and outflow properties of extended atmospheres (Parker, 1963). For the adiabatic atmospheres considered in this work, this structure is specified with the adiabatic index (Γ) whose calculation is described in the main text (§2.4). Here, we present raw model outputs for adiabatic calculations and describe the derivation of the adiabatic index via an example. Results of two sample adiabatic calculations (from Fig. 1 and 2 in the text) are shown in Table C1. Across four orders-of-magnitude of pressure variation, adiabatic variations in molecular weight (mainly due to $H_2$ recombination) and temperature are shown. By plotting log $P/P_0$ versus log $\rho/\rho_0$, we observe an approximately linear trend whose best-fit slope – which we derive using ordinary least squares regression – gives the adiabatic index due to the adiabatic relation ($P \sim \rho^{\Gamma}$).

To illustrate the importance of vertical structure in hydrodynamic outflows, we present the results of a sensitivity analysis of the vigor of outflow to small variations in the adiabatic index (Γ) (Table C2). For the baseline case, we consider one of the adiabatic solutions shown in Figure 4 of the main text ($\lambda_0$=11.95 and Γ=1.162, see Table 2) and arbitrarily vary the adiabatic index by a few percent in either direction. Small changes to the adiabatic index can alter the atmospheric structure from a hydrostatic structure with no loss to outflowing atmospheres experiencing prompt and large scale evacuation to space (Table C2). Departures of the thermal structure (Γ) from the adiabatic value ($\Gamma_{ad}$) can arise due to radiative coupling with the hot magma layer or to radiative losses due to infrared-active molecules. Full radiation-hydrodynamic simulations may enable more precise evaluation of the thermal structure of extended proto-lunar disk atmospheres.

**Table C1** – Variations in molecular speciation and temperature for adiabatic C-O-H atmospheres. $P_T$ – total pressure of the atmospheric level. $\mu$ – mean molecular weight. $T$ – atmospheric temperature. $\rho$ – gas density.

| System | $P_T$ (bars) | $\mu$ (amu) | T (K) | ρ ($kg/m^3$) |
|---|---|---|---|---|
| Earth | 196 | 22.1 | 4981 | 10.4 |
| | 19.6 | 23.4 | 3434 | 1.59 |
| | 1.96 | 23.8 | 2150 | 0.260 |
| | 0.196 | 23.9 | 1236 | 0.045 |
| | 0.0196 | 23.9 | 713 | 0.008 |
| Disk | 0.042 | 5.42 | 3005 | 9.01E-4 |
| | 0.0042 | 5.90 | 2557 | 1.15E-4 |
| | 4.2E-4 | 6.36 | 2202 | 1.44E-5 |
| | 4.2E-5 | 6.79 | 1890 | 1.79E-6 |
| | 4.2E-6 | 7.11 | 1515 | 2.35E-7 |

The adopted surface density for all the calculated disk models is $\sigma_T = 5 \times 10^7 kg/m^2$

In all cases the total system volatile mass fractions are [H] = 100 ppm, [C] = 100 ppm.

Proto-lunar disk calculations conducted near the Roche radius at 3 Earth radii.

ΔIW = -2. These calculated adiabats are those shown in Figures 1 and 2.

**Table C2** – Sensitivity analysis for outflows of C-O-H-Na disk atmospheres to variations in adiabatic index. Percent-level variations in the adiabatic index result in large variations in the outflow velocities and therefore mass-loss rates. $r_0$ – wind base radius. $f_{Na}^V$ – fraction of Na inventory vaporized. $\Gamma$ – adiabatic index. $\lambda_0$ – base escape parameter. $v_0$ – base outflow velocity. $a_0$ – base isothermal sound speed. $r_c$ – critical radius. $M_0$ – base Mach number.

| $r_0$ ($R_E$) | $f_{Na}^V$ | $\Gamma$ | $\lambda_0$ | $x_0$ | $w_0$ | $v_0$ (m/s) | $a_0$ (m/s) | $r_c$ ($R_E$) | $M_0$ |
|---|---|---|---|---|---|---|---|---|---|
| 3 | 0.50 | 1.192‡ | 11.95 | 0.011 | 0.002 | 0.67 | 1320 | 273 | 4E-5 |
| 3 | 0.50 | 1.182‡ | 11.95 | 0.022 | 0.011 | 5.23 | 1320 | 136 | 4E-3 |
| 3 | 0.50 | 1.172‡ | 11.95 | 0.034 | 0.027 | 15.8 | 1320 | 88 | 0.011 |
| 3 | 0.50 | 1.162† | 11.95 | 0.046 | 0.046 | 31.0 | 1321 | 65.93 | 0.022 |
| 3 | 0.50 | 1.152‡ | 11.95 | 0.057 | 0.068 | 50.8 | 1320 | 52.6 | 0.036 |
| 3 | 0.50 | 1.142‡ | 11.95 | 0.068 | 0.091 | 73.9 | 1320 | 44.1 | 0.052 |
| 3 | 0.50 | 1.132‡ | 11.95 | 0.079 | 0.114 | 99.2 | 1320 | 38.0 | 0.071 |

† Adiabatic index calculated using multi-species adiabats and the procedure described in §2.4.

‡ Sensitivity test with adiabatic index arbitrarily varied around the calculated value ($\Gamma$=1.162)

**Appendix D. Oxygen fugacity of the proto-lunar disk atmosphere at relevant thermal states**

The predominance of $H_2$ relative to $H_2O$ and H relative to OH is directly related to the oxygen fugacity prevailing in the proto-lunar disk, as can be verified by considering gas phase-equilibria (see Table B1). Since we have invoked an H-$H_2$-rich proto-lunar disk atmosphere, this prompts the crucial question of the oxygen fugacity prevailing in the proto-lunar disk at the tiem of onset of the volatile-element (C-O-H-Na) atmosphere. Early in the evolution, for example, immediately after the melt-vapor disk is produced by a giant impact, it is thought that the disk is substantially vaporized with estimates ranging from ≈20-80% vapor by mass depending on the nature of the Moon-forming event (Nakajima and Stevenson, 2014). Under such large degrees of vaporization, pure (metal-free) silicate models (Van Buchem et al., 2025; Visscher and Fegley, 2013) may be reasonably accurate, given that the Moon is – and presumably the proto-lunar disk was – composed of >98% silicate by mass (Williams et al., 2014). However, as the proto-lunar disk experiences secular cooling and silicate condensation, the vapor fraction of the disk decreases and the ≈1-2 wt% metallic Fe inventory now residing in the lunar core – which is assumed to have negligible solubility in silicate melt – becomes an increasingly important component in the disk atmosphere and begins to exert influence on the prevailing conditions. This influence has not been captured by previous models. Indeed, the presence of "excess" Fe (Fe unaccompanied by O) and the associated condensation Fe-rich metals links the distinct chemical conditions of "highly oxidizing" silicate vapor in the proto-lunar disk (Visscher and Fegley, 2013) to the inferred "highly reducing" character of lunar magmas (Wadhwa, 2008). Once metallic Fe has condensed into a separate liquid phase, $fO_2$ can be calculated in a straightforward manner as its value is then buffered by the iron-wüstite reaction. The small mass of metallic Fe relative to silicates is of no consequence so long as metallic Fe has condensed as a separate phase because oxygen fugacity is an intensive variable and is independent of the proportions of the phases present (Frost, 1991). These considerations lead to the conclusion that Fe vapor saturation and liquid metal condensation and oxygen fugacity evolution in the proto-lunar disk atmosphere are inseparable processes. Here, we consider the question of how much metallic Fe (in wt%) is required in the proto-lunar disk to saturate the atmosphere and sufficiently lower oxygen fugacity to enable condensation of metals. We consider the state in which metallic Fe has condensed as a separate liquid and calculate partial pressures of Fe vapor [Fe(liquid)⇔Fe(vapor)] characterizing the co-existing atmosphere. We use an equation of the form $P_{Fe}=P_0\exp(-T_0/T)$ with $P_0=2.3\times10^6$ bars and $T_0=4.6\times10^4$ K, which passes through the boiling point of Fe at 1 bar pressure (3134 K) to an accuracy of a fraction of a percent. To convert partial pressure to surface density, the mean molecular weight of the atmosphere is needed (e.g., see Equation E8). For the purposes of illustration, we consider pure Fe vapor atmospheres in converting from partial pressure to surface density and do not attempt to explicitly capture the complex interplay between Fe vapor and silicate vapor and volatile element-rich (C-O-H-Na) atmospheres. The critical metallic Fe mass fraction needed to saturate the disk atmosphere in Fe vapor, overcome the oxidizing influence of silicate vapors, and to establish a chemically reducing atmosphere is given in Table D2. For the 1-2 wt% lunar metal inventory (Williams et al., 2014), metallic Fe condensation and the transition to a chemically reducing C-O-H-Na atmosphere occurs at midplane temperatures near $T_M$=3,000 K (see Table D2). This result need not be overinterpreted: even atmospheres with subsaturated midplane regions can condense metallic Fe clouds at altitude, which scavenge oxygen and thus generate $H$-$H_2$-Na-rich *upper* disk atmospheres. This calculation illustrates the influence of metallic Fe on the $fO_2$ evolution of the proto-lunar disk. We conclude that metals, although constituting a small fraction of the lunar mass, played a critical role in the evolution of the proto-lunar disk by producing low molecular weight gas compositions susceptible to outflows.

**Table D1** – Oxygen fugacity for metal-free (LavAtmos) and metal-saturated (this work) magmatic systems. All values evaluated at 3,000 K. The similarity of the volatile-free and volatile-bearing LavAtmos $fO_2$ values reflects the small buffering capacity of volatile gases relative to the liquids at this temperature.

| Model | $P_{O2}$ (bars) | $P_{H2}/P_{H2O}$ |
|---|---|---|
| LavAtmos (volatile-free) | 0.0538 | 0.197 |
| LavAtmos (BSE H and C) † | 0.0535 | 0.197 |
| Metal-saturated (IW-1) | 1.64E-4 | 3.56 |
| Metal-saturated (IW-2) | 1.64E-5 | 11.3 |

†Input parameters are volatile pressure = 0.021 bars, with volatile composition consisting of 92.3% H and 7.7% C (H/C=12, as inferred for the BSE by Dasgupta and Hirschmann, 2009).

**Table D2** – Mass of metallic Fe needed to saturate the vapor and condense metallic droplets throughout the column. T – midplane temperature. $P_{Fe}$ – saturation vapor pressure of Fe. $a_0$ – isothermal sound speed. $g_{eff}$ – effective gravity in the disk atmosphere. $\sigma_v^{Fe}$ – surface density of Fe vapor in a saturated disk. $[Fe]_{crit}$ – critical Fe mass fraction for liquid metal condensation.

| $T_M$ (K) | $P_{Fe}$ (bars) | $a_0$ (m/s) | $g_{eff}$ (m/s$^2$) | $\sigma_{Fe}^v$ (kg/m$^2$) | $[Fe]_{crit}$ (wt%) |
|---|---|---|---|---|---|
| 2500 | 0.024 | 607 | 0.058 | 4.0E4 | 0.08 |
| 2750 | 0.125 | 637 | 0.061 | 2.1E5 | 0.41 |
| 3000 | 0.504 | 665 | 0.064 | 7.9E5 | 1.58 |
| 3250 | 1.640 | 693 | 0.066 | 2.5E6 | 4.95 |
| 3500 | 4.507 | 719 | 0.069 | 6.6E6 | 13.1 |

The adopted surface density for all disk calculations here is $\sigma_T = 5 \times 10^7 kg/m^2$

†Effective gravity is for the disk is calculated using $g_{eff} = a_0\Omega/\sqrt{2\pi}$

Disk calculations conducted near Roche radius at 3 Earth radii ($\Omega$ = 2.4x10$^{-4}s^{-1}$)

## Appendix E. Procedure for calculating transonic disk wind solutions

The adopted procedure for deriving equations for adiabatic rotating disk winds is closely analogous to those describing adiabatic non-rotating spherical winds. The latter has been extensively described in the context of stellar outflows (Lamers and Cassinelli, 1999). Here, we develop the analogous equations for describing the former. The term $\rho^{-1}dP/dr$ in the momentum equation can be eliminated by assuming stationary flow ($\dot{M} = constant$), differentiating the log of Equation 6 in the main text and using an ideal gas equation of state ($P = \rho kT/\mu$):

$$\frac{1}{\rho}\frac{dP}{dr} = \Gamma\frac{\mathrm{P}}{\rho}\frac{dln\rho}{dr} = \Gamma\frac{kT}{\mu}\frac{dln\rho}{dr} = \Gamma\frac{kT}{\mu}\left(-\frac{1}{v}\frac{dv}{dr} - \frac{2}{r}\right) \quad \text{(E1)}$$

where we use the polytropic definition ($dlnP = \Gamma \mathrm{dln}\rho$). The momentum equation is:

$$v\frac{dv}{dr} + \Gamma\frac{kT}{\mu}\left(-\frac{1}{v}\frac{dv}{dr} - \frac{2}{r}\right) + \frac{GM}{r^2}\left(1 - \frac{r_0}{r}\right) = 0 \quad \text{(E2)}$$

or, by collecting the velocity derivatives:

$$\frac{1}{v}\frac{dv}{dr} = \left(\frac{2\Gamma a^2}{r} - \frac{GM}{r^2}\left[1 - \frac{r_0}{r}\right]\right)/(v^2 - \Gamma a^2) \quad \text{(E3)}$$

where a is the isothermal sound speed ($a = \sqrt{kT/\mu}$), which varies with r because T and μ vary with r (see Fig. 1 and 2). Equation E3 describes flow velocity as a function of radius for a family of hydrodynamic solutions. Of these, only one – the *transonic* solution – is subsonic at the base, passes through a *critical* point, and becomes supersonic at large radii. This is the sought-after transonic solution for a Parker wind. The critical point, through which the transonic solution passes, is a singularity in Equation E3 in which both the numerator and denominator of the right hand side simultaneously vanish, from which emerge the constraints:

$$2\Gamma a_c^2 = \frac{GM}{r_c}\left(1 - \frac{r_0}{r_c}\right) \quad \text{(E4)}$$

and

$$v_c^2 = \Gamma a_c^2 = \frac{GM}{2r_c}\left(1 - \frac{r_0}{r_c}\right) \quad \text{(E5)}$$

where $v_c$, $a_c$, and $r_c$ are the flow velocity, isothermal sound speed, and radial distance, all evaluated at the critical point. Equation E5 relates the critical velocity ($v_c$) to the critical radius ($r_c$). To solve for these two quantities and thus characterize outflow solutions for a given set of input parameters, another constraint is needed and provided by the energy equation. The momentum equation (Equation 5 in main text) can be written in alternate form by making use of the polytropic relation $[P \propto (P/\rho)^{\Gamma/\Gamma-1}]$, which is another statement of the more familiar definition ($P \propto \rho^{\Gamma}$). The term $\rho^{-1} dP/dr$ can be written:

$$\frac{1}{\rho}\frac{dP}{dr} = \frac{\mathrm{P}}{\rho}\frac{dlnP}{dr} = \frac{\Gamma}{\Gamma-1}\frac{\mathrm{P}}{\rho}\frac{dln(P/\rho)}{dr} = \frac{\Gamma}{\Gamma-1}\frac{d(P/\rho)}{dr} \tag{E6}$$

which holds for $\Gamma \neq 1$. Using Equation S6, the momentum equation is:

$$v\frac{dv}{dr} + \frac{\Gamma}{\Gamma-1}\frac{d}{dr}\left(\frac{P}{\rho}\right) + \frac{GM}{r^2}\left(1 - \frac{r_0}{r}\right) = 0 \tag{E7}$$

which, with an ideal gas equation of state, can be integrated to:

$$\frac{d}{dr}\left(\frac{v^2}{2} + \frac{\Gamma}{\Gamma-1}\frac{kT}{\mu} - \frac{GM}{r}\left[1 - \frac{r_0}{2r}\right]\right) = 0 \tag{E8}$$

Equation E8 indicates that there is a constant along the flow in a polytropic disk wind that has units of specific energy:

$$e_\Gamma \equiv \frac{v^2}{2} + \left(\frac{\Gamma}{\Gamma-1}\right)\frac{kT}{\mu} - \frac{GM}{r}\left[1 - \frac{r_0}{2r}\right] \tag{E9}$$

where $e_\Gamma$ is sometimes called 'the energy constant' (Lamers and Cassinelli, 1999) or 'the Bernoulli parameter' (Gammie et al., 2016). It is the kinetic energy of the gas at infinity and must be positive for an outflow to arise. Using E5 and E9, the energy constant at the critical point is:

$$e_{\Gamma,\mathrm{c}} = \frac{v_c^2}{2} + \frac{v_c^2}{\Gamma-1} - 2v_c^2\left(1 - \frac{r_0}{2r_c}\right)/\left(1 - \frac{r_0}{r_c}\right) \tag{E10}$$

while at the lower boundary of the wind, it is:

$$e_{\Gamma,0} = \frac{v_0^2}{2} + \frac{v_c^2}{\Gamma-1}\left(\frac{v_0}{v_c}\right)^{1-\Gamma}\left(\frac{r_0}{\mathrm{r}_c}\right)^{2-2\Gamma} - v_c^2\frac{r_c}{r_0}\frac{1}{(1-\frac{r_0}{r_c})} \tag{E11}$$

where we use $a_0^2 = a_c^2(\mathrm{T}_0/\mathrm{T}_c)(\mu_c/\mu_0)$ to scale the isothermal sound speed, and the polytropic relation $(T/\mu) \propto (\rho/\rho_c)^{\Gamma-1}$ and spherical continuity equation $(\rho \propto v^{-1}r^{-2})$ and the relations in E5 to express all variables relative to critical point values. In order for lower boundary conditions $(r_0, v_0)$ to connect to critical point conditions $(r_c, v_c)$, the two must be characterized by the same energy constant:

$$e_{\Gamma,c} = e_{\Gamma,0} \tag{E12}$$

E5 and E10-E12 can be expressed using dimensionless variables $x_0(\equiv r_0/r_c)$ and $w_0(\equiv v_0/v_c)$:

$$\frac{1}{2} + \frac{1}{\Gamma-1} - 2\left(1 - \frac{x_0}{2}\right)/(1 - x_0) = \frac{w_0^2}{2} + \frac{1}{\Gamma-1}{w_0}^{1-\Gamma}{x_0}^{2-2\Gamma} - \frac{1}{x_0(1-x_0)} \tag{E13}$$

$${w_0}^{\Gamma-1}{x_0}^{2\Gamma-3} = \frac{\lambda_0}{2\Gamma}(1 - x_0) \tag{E14}$$

where, in E14, we have used $a_0^2 = a_c^2(\mathrm{T}_0/\mathrm{T}_c)(\mu_c/\mu_0)$, the polytropic relation $(T/\mu) \propto (\rho/\rho_c)^{\Gamma-1}$ and the spherical continuity equation $(\rho \propto v^{-1}r^{-2})$ to express all variables in terms of the basal quantities. E13 and E14 can be used to simultaneously solve for critical quantities characterizing transonic wind solutions ($x_0$, $w_0$) for any disk atmosphere described by the Jeans escape parameter evaluated at the base in/near the midplane ($\lambda_0$) and adiabatic index characterizing vertical structure ($\Gamma$). Of course, physical solutions for disk outflows only exist for atmospheres that are hydrodynamic, that is, those in which $\frac{\Gamma-1}{\Gamma}\lambda_0 < 2$ (see Equation 4 in the main text). Using these equations, we simultaneously solve for the basal flow velocity ($w_0$) and the basal radius ($x_0$) for outflowing atmospheres. The reader can verify that the hydrodynamic disk wind solutions given in Table 2 simultaneously satisfy E13 and E14.

It may be necessary or useful to express these quantities in terms of physical units. By definition:

$$r_c = r_0/x_0 \tag{E15}$$

where $r_0$ is the radius of the wind base defined by the angular momentum of the escaping gas (see Equation 5 in main text). For the critical velocity, we use the first equality in E5:

$$v_c = \sqrt{\frac{\Gamma k T_c}{\mu_c}} = \sqrt{\frac{\Gamma k T_0}{\mu_0}} w_0^{(\Gamma-1)/2} x_0^{\Gamma-1} \quad \text{(E16)}$$

where we have used $a_0^2 = a_c^2(T_0/T_c)(\mu_c/\mu_0)$, the polytropic relation $(T/\mu) \propto (\rho/\rho_c)^{\Gamma-1}$ and the spherical continuity equation ($\rho \propto v^{-1}r^{-2}$) to relate the temperature and mean molecular weight at the base, which are input parameters to the escape model, to those at the critical point. The basal velocity, which enters into expressions for the mass loss rate, is given by:

$$v_0 = w_0 v_c = \sqrt{\frac{\Gamma k T_0}{\mu_0}} w_0^{(\Gamma+1)/2} x_0^{\Gamma-1} \quad \text{(E17)}$$

Finally, it may be useful to express the basal Mach number, which is the basal velocity relative to the local sound speed, because this is the governing parameter that influences the magnitude of kinetically-controlled evaporative isotopic fractionation for an atmosphere that experiences liquid-vapor exchange at/near the base while undergoing hydrodynamic escape from the top (Pritchard and Stevenson, 2000). The basal Mach number is given purely by the dimensionless variables:

$$M_0 = \frac{v_0}{\sqrt{\Gamma} a_0} = w_0^{(\Gamma+1)/2} x_0^{\Gamma-1} \quad \text{(E18)}$$

and is shown, for the hydrodynamic disk atmospheres considered here, in Figure 4. The outflow properties of all atmospheres considered in this work are calculated using input parameters ($\lambda_0$, $\Gamma$) (Table 2) in E13-E15 and E17-E18.

**Appendix F. Negligible D/H fractionation accompanying hydrodynamic disk escape**

The question of lunar mantle D/H is relevant if it can constrain the proposed model of lunar volatile depletion. We consider three processes that discriminate D from H: magma-vapor equilibration, kinetically-controlled degassing, and mass-fractionation during hydrodynamic escape. First, equilibrium isotope fractionation between magma and vapor has previously been calculated and shown to be < 10 per mil for conditions prevailing in the proto-lunar disk (Pahlevan et al., 2016). This fractionation is insignificant on the scale of the precision of the inferences of D/H in lunar interior reservoirs. Second, there is a possibility of kinetic isotope fractionation accompanying evaporation into a subsaturated medium. Indeed, this is the process that has recently been invoked on the basis of moderately volatile element isotopes (e.g., Zn, Rb, K, etc.) (Dauphas et al., 2022; Nie and Dauphas, 2019), which we propose may be a signature of hydrodynamic outflows (§3.4). For kinetic degassing of H, we estimate D/H fractionation using Equation 3 of (Nie and Dauphas, 2019):

$$\delta_L - \delta_0 \approx (1 - S_i)\Delta_{kin,i}^{V-L} ln f_i \tag{F1}$$

Here, $\delta_L$ is the part per thousand offset of D/H from a reference composition after distillation, $\delta_0$ is the initial D/H composition of the reservoir before distillation (which we take to be equal zero), $S_i$ the degree of subsaturation of the vapor into which the liquid is evaporating or degassing, which is ≈0.99 (Nie and Dauphas, 2019) based on the moderately volatile elements, $\Delta_{kin,i}^{V-L}$ is the part per thousand kinetic isotopic fractionation between liquid and degassing vapor. In the absence of experimental data characterizing the mass-dependence of kinetic degassing of H from magmas, we use experimental data from potassium evaporation (Richter et al., 2011):

$$\Delta_{kin,i}^{V-L} = [(m_1/m_2)^{0.43} - 1] \times 10^3 \tag{F2}$$

where $m_1$ and $m_2$ are the mass of the isotopically light and heavy degassing isotopologues. In the proto-lunar disk, H is expected to dissolve in silicate melts mainly as OH and is expected to exsolve from magmas as water vapor, irrespective of its equilibrium gas phase speciation (mainly H and $H_2$ in our models). Accordingly, our expectation is that $m_1$ and $m_2$ are 18 and 19 proton masses, respectively, in which case, $\Delta_{kin,i}^{V-L} = -23‰$. Finally, $f_i$ represents the fraction of the element remaining after the distillation sequence, for which we take a range (0.1-0.01) to illustrate the range of expected fractionations. Substituting these factors into Equation F1, we calculate δD for the residual dissolved proto-lunar hydrogen to be in the range 0.5-1‰, mainly reflecting the fact that H degasses as $H_2O$ and is therefore fractionated to an extent comparable in magnitude to that of potassium (Wang and Jacobsen, 2016). Even taking the extreme and unproven assumption that dissolved OH kinetically degasses from silicate melts in the proto-lunar disk as atomic H (with $m_1$ and $m_2$ equal to 1 and 2 proton masses, respectively), the residual dissolved δD is 6-12‰, and below the threshold of detection for lunar sample studies (Saal et al., 2013). The reason for a relatively small fractionation is the near-equilibrium state characterizing lunar volatile depletion (S≈0.99 where S=1 is thermodynamic equilibrium) (Nie and Dauphas, 2019), the same circumstance that delayed detection of fractionated lunar K until (Wang and Jacobsen, 2016). Finally, there is the possibility of mass-fractionation accompanying hydrodynamic escape itself, the process thought to have fractionated terrestrial Xe (Zahnle et al., 2019). Because such a diffusively-mediated process is effective for relatively "gentle" flows that occur over millions of years, such a mass-fractionation process is not expected to be significant for the vigorous outflows occurring on a timescale of weeks we consider here. Using the formalism of (Hunten et al., 1987) and considering only the flux of $H_2$, we estimate the crossover mass below which gaseous species are entrained for an outflow in Table 2 ($r_0$ = 3 $R_E$, $T_M$=3,000 K, log$fO_2$=IW-2 and $f_{Na}^V$=0.5) to be ≈$10^{10}$ amu. This substantiates modeling the H-$H_2$-Na-rich outflow as a single fluid with negligible mass separation. We conclude that D/H fractionation during the proposed loss episode was negligible and other geochemical indicators (e.g., high H/C, see Table 1) may be better suited for identifying residual primordial water in lunar samples.

**CRediT authorship contribution statement**

**Kaveh Pahlevan:** Writing – original draft, Writing – review & editing, Investigation, Visualization, Formal analysis. **Andrew Youdin:** Writing – review & editing, Investigation, Visualization. **Paolo A. Sossi:** Writing – review & editing, Formal analysis, Investigation.

**Competing Interest Declaration.**

The authors declare that they have no competing financial interests or personal relationships that could have appeared to influence the work reported in this paper.

**Acknowledgements**

The authors acknowledge valuable discussions with Miki Nakajima, Steven Desch, Jeff Taylor, Sebastian Charnoz, and Dave Stevenson and comments on an earlier draft by Miki Nakajima that greatly helped to improve the manuscript. K.P. acknowledges support from NASA (80NSSC20K0584).

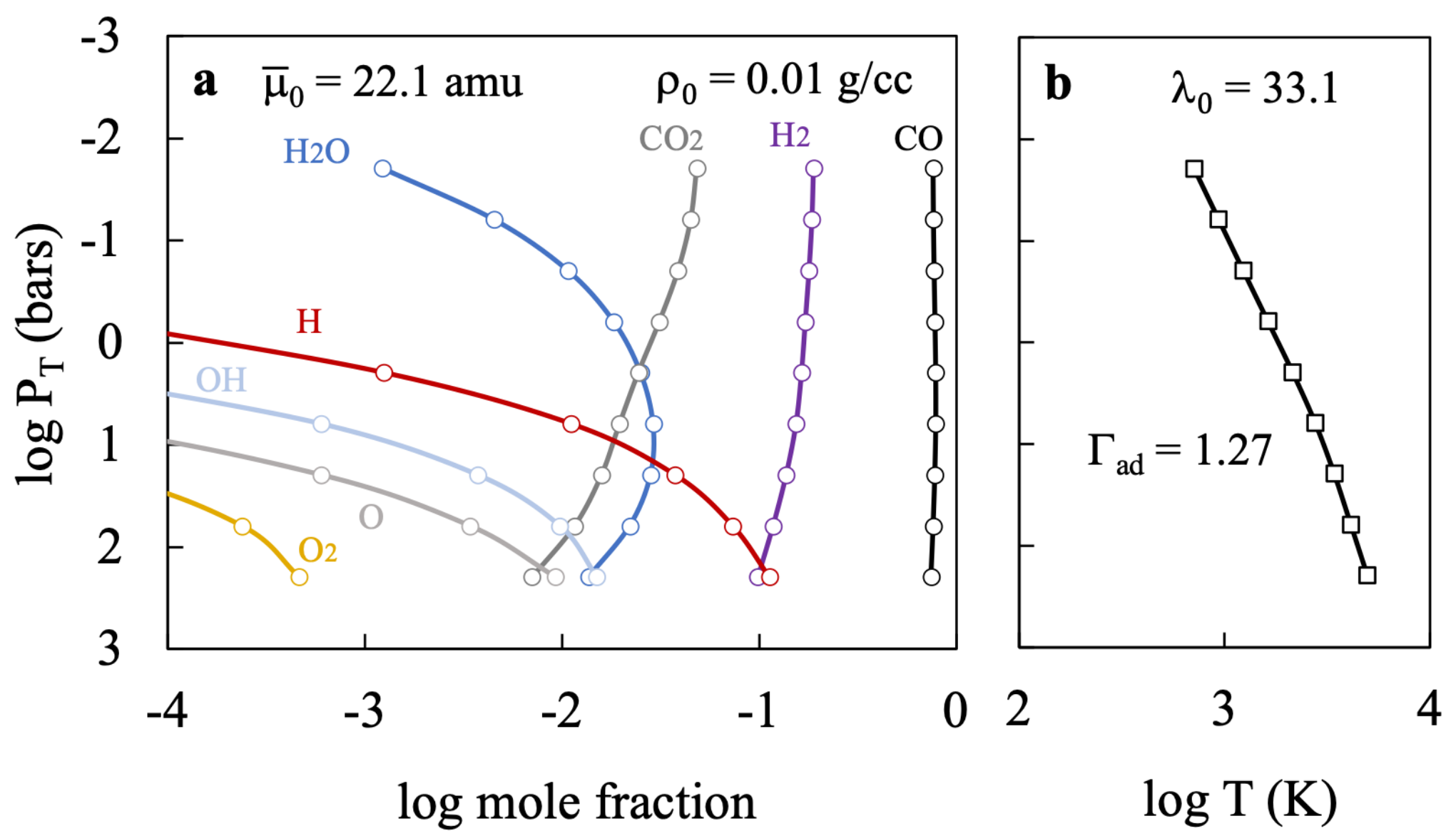

**Figure 1 – Composition and adiabatic structure of Earth's post-giant impact atmosphere a.** Mole fraction ($\equiv P_i/P_T$) of the dominant C-O-H species versus atmospheric depth in units of total pressure ($P_T$). Chemical composition of the atmosphere is determined via equilibrium partitioning with the magma ocean with the conditions of equilibrium shown at the atmospheric base (the highest pressure level plotted), whereas vertical variations in temperature and speciation are calculated using a reactive adiabatic model. Modeled terrestrial atmospheres are dominated by CO at all altitudes, resulting in a compact high molecular weight atmosphere that is strongly gravitationally bound. **b.** The predominance of CO – which does not thermally dissociate in this environment – produces a largely non-reactive atmosphere in which the adiabatic index $\Gamma \equiv \left(\frac{dlnP}{dln\rho}\right)_S$ is nearly equal to that of an inert high-temperature gas (Γ≈1.3). Such large values of the adiabatic index result in thermal structures with cooler upper atmospheres produced by adiabatic decompression, a feature that decreases the scale height at altitude and stabilizes the atmosphere against outflow. Plotted atmospheric composition is set via equilibration at the surface of a magma ocean near T=5,000 K and log*f*$O_2$=IW-2 (see Table 1 for further details on model atmospheres).

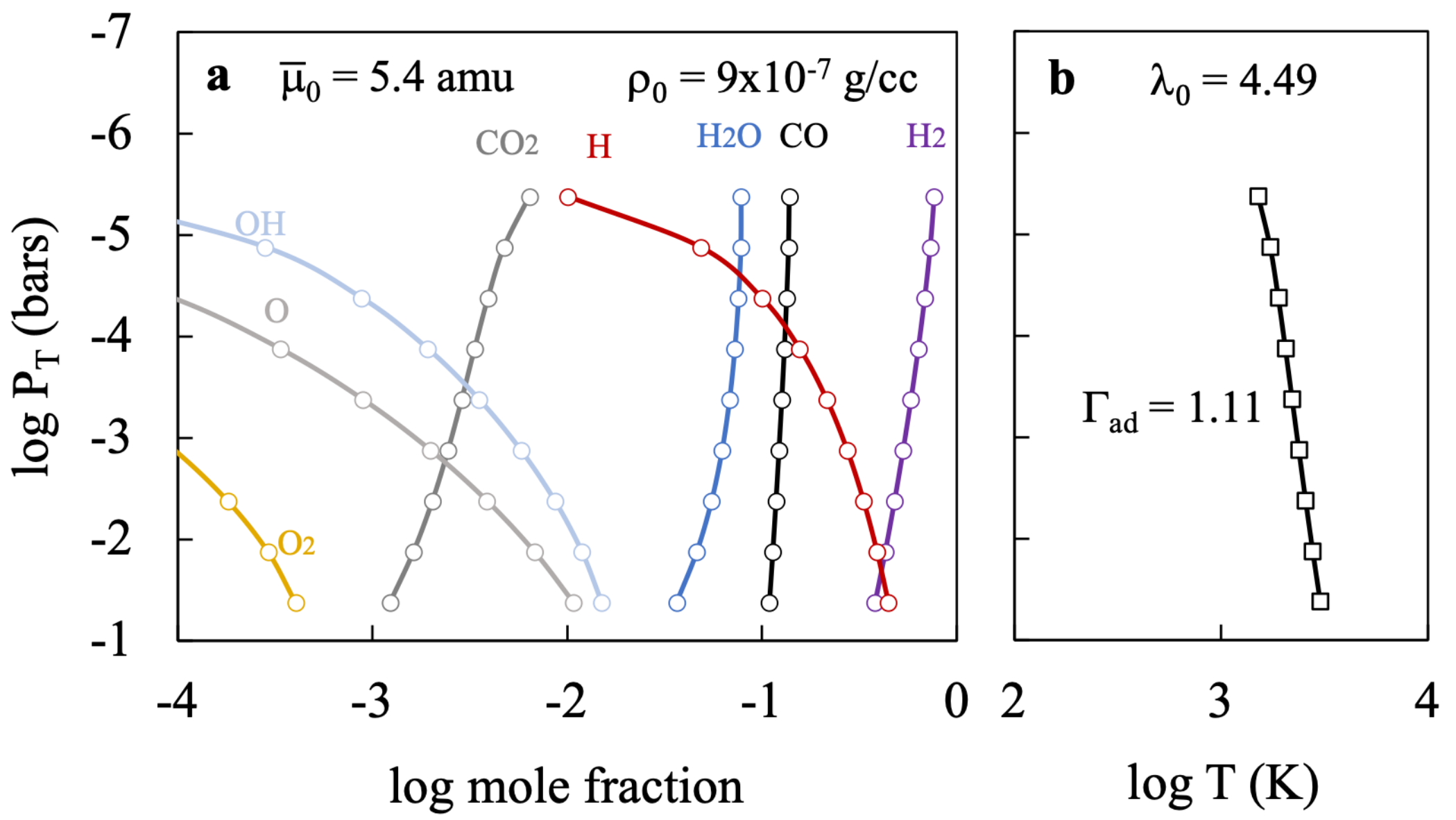

**Figure 2 – Composition and adiabatic structure of a proto-lunar disk atmosphere a.** The same quantities are plotted as in Figure 1, but the atmosphere is confined by a gravity field at 3 Earth radii near the Roche radius. Due to the low effective gravity and shallow depth of the magma layer, both H and C are nearly fully outgassed into the disk atmosphere (Table 1). The H/C atomic ratio of the disk atmosphere thus mainly reflects the H/C atomic ratio of the bulk disk, i.e., ≈12 for a BSE-composition inventory. In equilibrium with a metal-saturated magma layer, the major gas species are H-$H_2$ and CO with $H_2O$ and $CO_2$ present as minor species. The predominance of H-$H_2$ lowers the mean molecular weight, which produces an inflated atmosphere more susceptible to hydrodynamic outflow. **b.** Atmospheres are largely atomic at depth but more molecular at altitude due to recombination driven by adiabatic expansion and cooling. Equilibrium recombination (2H→$H_2$) releases large quantities of heat throughout the disk atmosphere, effectively transporting magmatic heat to high altitudes and producing nearly isothermal structures ($\Gamma_{ad}$=1.11 where Γ=1 is isothermal). Plotted atmospheric composition is set via equilibration with a metal-saturated magma layer near T=3,000 K and log$f$O$_2$=IW-2 (see Table 1 for details of model atmospheres).

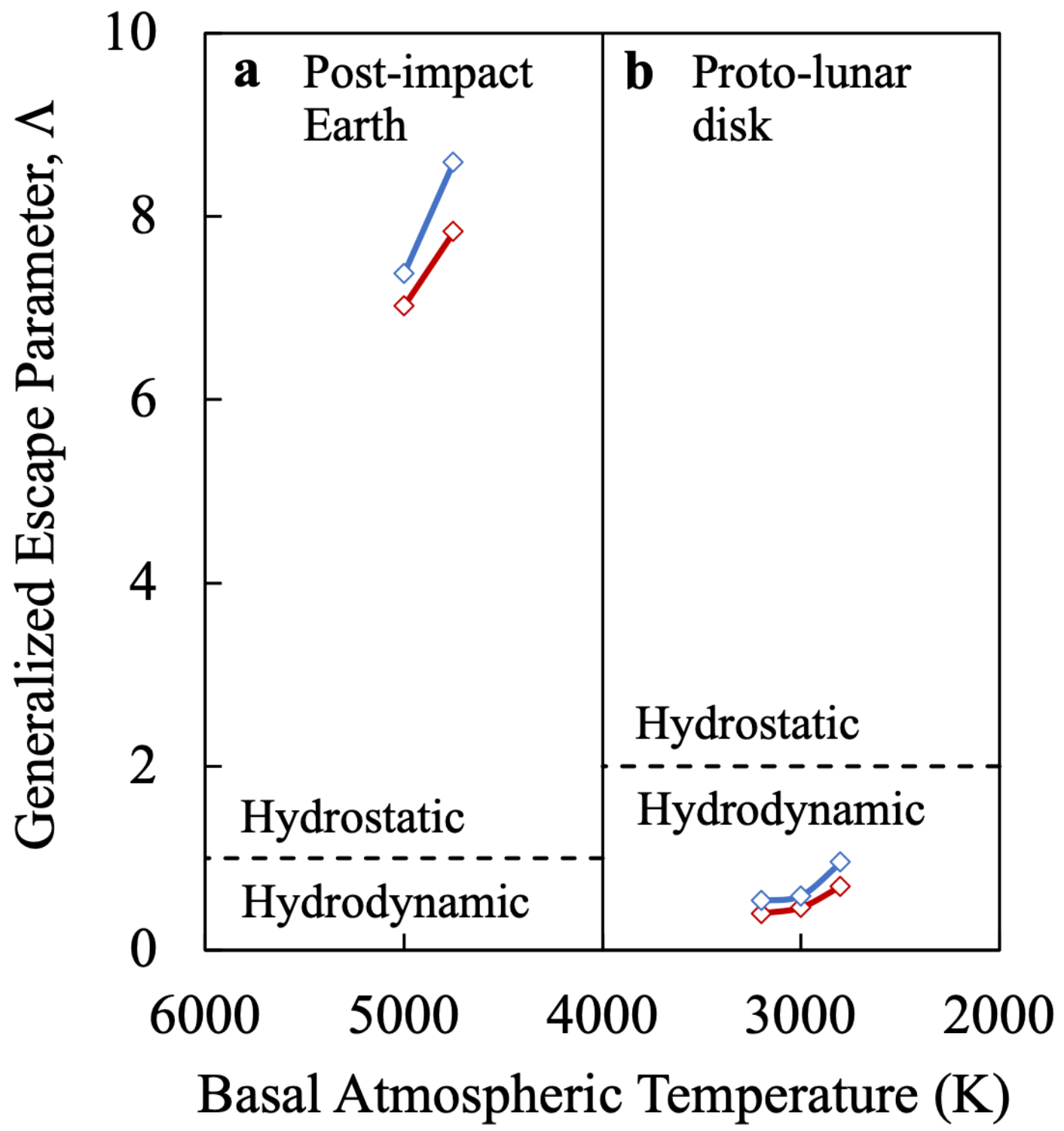

**Figure 3 – Hydrodynamic stability of adiabatic post-giant impact C-O-H atmospheres. a.** The volatile-rich atmosphere of Earth is in the regime of hydrostatic equilibrium and a gravitationally-bound reservoir. High mean molecular weight ($\bar{\mu} \gtrsim 22$ amu), a steep thermal gradient ($\Gamma_{ad}$=1.27), and a deep position in Earth's potential well combine to produce a compact structure far from the outflow condition. **b.** The volatile-rich atmosphere of the proto-lunar disk – evaluated near the classical Roche radius at 3 Earth radii – is unable to maintain hydrostatic equilibrium and is unstable with respect to rapid hydrodynamic expansion out into interplanetary space. Low mean molecular weight ($\bar{\mu} \lesssim$ 5-6 amu), a nearly isothermal structure ($\Gamma_{ad}$=1.11), a shallow position in Earth's potential well, and pre-existing orbital motion ($\Lambda_{crit}$=2) combine to produce a spatially extended and expanding structure. The two colors correspond to atmosphere compositions determined by equilibration with underlying liquids at IW-1 (blue) and IW-2 (red). See §2.5 for a quantitative discussion of the adopted criteria for hydrodynamic onset.

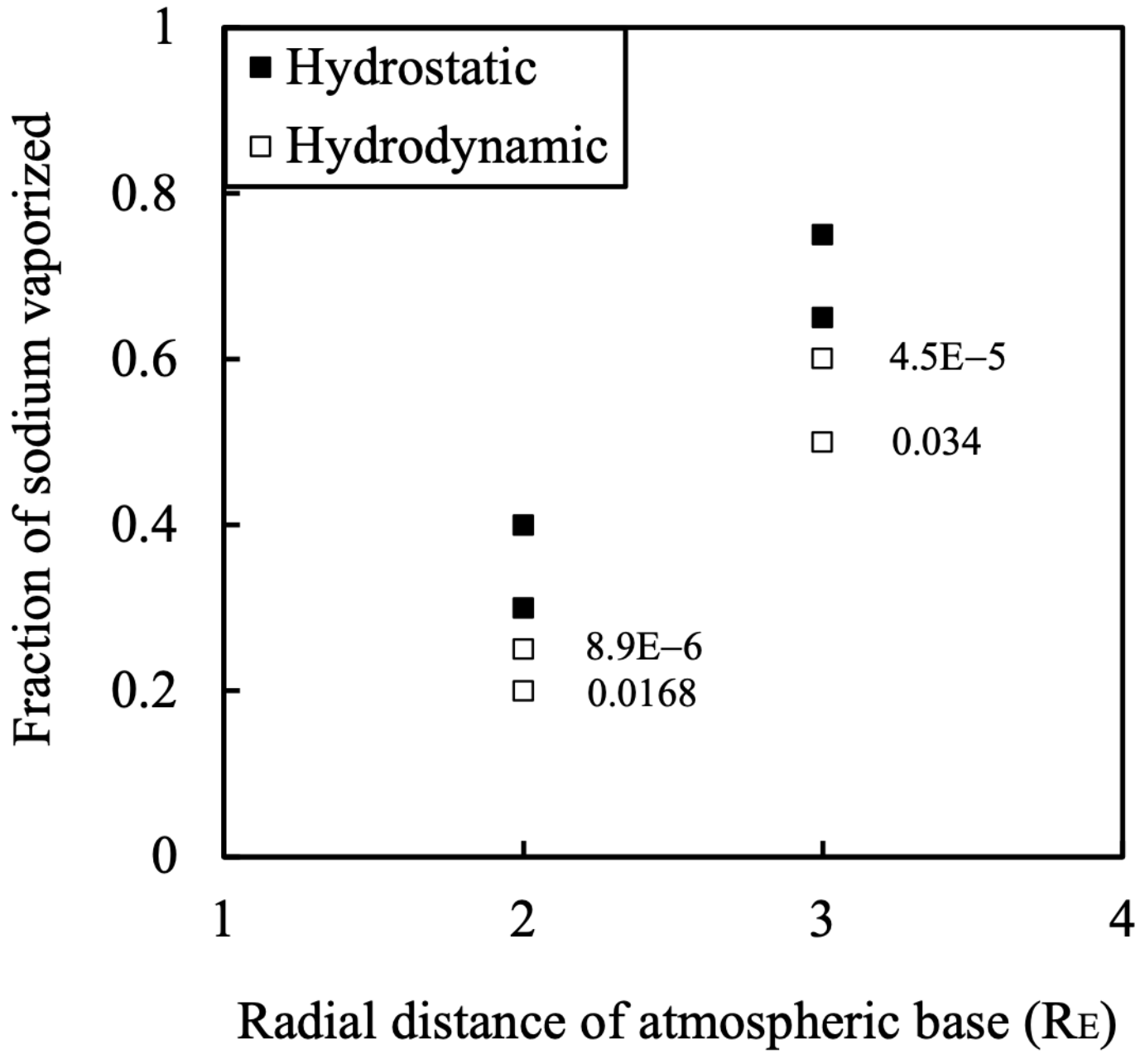

**Figure 4 – Dependence of entrained Na abundance on radial distance of the wind base.** When sodium in the Roche-interior (< $3R_E$) proto-lunar disk is fully vaporized, the atmosphere has a high molecular weight and a relatively compact structure and is hydrostatic. As Na partially condenses into the disk liquid, the mean molecular weight of the remnant volatile-rich atmosphere decreases and the atmosphere vertically expands. At a critical sodium abundance, the atmosphere crosses the threshold of hydrostatic breakdown, and further sodium condensation and associated atmospheric inflation leads to rapidly increasing mass outflow rates. Labels for hydrodynamic data points refer to the midplane Mach number, which corresponds to a mass loss rate. For disk material at 3 $R_E$ (2 $R_E$), the abrupt transition to rapid removal occurs at 60% (25%) vaporized sodium for a BSE-composition disk. These adiabatic solutions represent the onset of winds at a critical Na abundance and suggest that substantial MVE inventories can be entrained in outflows launched from a Roche-interior proto-lunar disk. Atmospheric composition here is set via equilibration with underlying liquids near T=3,000 K and $\log fO_2$=IW-2 with total disk surface density and hydrogen mass fraction of $\sigma = 5\text{x}10^7$ kg/m$^2$ and [H] = 100 ppm, respectively (see Table 2 for details on outflows).

**Table 1** – Partitioning of C-O-H during magma ocean-atmosphere equilibration. $T_0$ – equilibration temperature. ΔIW – offset of log$fO_2$ from iron-wüstite buffer. $P_T$ – equilibration pressure. $x_H^L$ – hydrogen mass fraction of silicate melt. $x_C^L$ – carbon mass fraction of silicate melt. $m_0$ – basal mean molecular weight. $g_{eff}$ – effective gravity. $H_p$ – pressure scale height. See Appendix B for the model thermodynamics and mechanics that relate these quantities.

| | $T_0$ (K) | ΔIW | $P_T$ (bars) | $x_H^L$ (ppm) | $x_C^L$ (ppm) | $\mu_0$ (amu) | $g_{eff}$ † ($m/s^2$) | $H_p$* (km) |
|---|---|---|---|---|---|---|---|---|
| Earth | 4980 | -2 | 196 | 96.1 | 0.21 | 22.1 | 10.0 | 186 |
| | 5019 | -1 | 202 | 98.1 | 0.72 | 25.0 | 10.0 | 166 |
| | 4757 | -2 | 195 | 96.4 | 0.21 | 22.7 | 10.0 | 173 |
| | 4740 | -1 | 198 | 98.4 | 0.75 | 25.7 | 10.0 | 152 |
| Disk | 3005 | -2 | 0.042 | 2.3 | 8E-6 | 5.4 | 0.205 | 8920 |
| | 3002 | -1 | 0.048 | 4.1 | 3E-5 | 7.3 | 0.176 | 7660 |
| | 2810 | -2 | 0.038 | 2.5 | 8E-6 | 5.9 | 0.190 | 8270 |
| | 2802 | -1 | 0.042 | 4.3 | 3E-5 | 7.6 | 0.167 | 7270 |

The surface density for all the calculated disk models is $\sigma_T = 5 \times 10^7 kg/m^2$

In all cases the total system volatile mass fractions are [H] = 100 ppm, [C] = 100 ppm.

Proto-lunar disk calculations conducted near the Roche radius at 3 Earth radii.

†Effective gravity is constant for the Earth and given by $g_{eff} = a_0\Omega/\sqrt{2\pi}$ for the disk

*Scale height calculated with $H_p = kT_0\ /\mu_0 g$ for Earth and $H_p = \sqrt{2}a_0\ /\Omega$ for the disk

**Table 2** – Flow properties of C-O-H-Na disk atmospheres as a function of base radius ($r_0$) and the fraction of the sodium inventory vaporized ($f_{Na}^V$). $\Gamma$ – adiabatic index. $\lambda_0$ – base escape parameter. $x_0$ – base radius in units of critical radii. $w_0$ – base flow velocity in units of critical velocity. $v_0$ – base flow velocity. $a_0$ – base isothermal sound speed. $r_c$ – critical radius. $M_0$ – base Mach number. See the escape model in the Appendix E for definitions and relations between these quantities.

| $r_0$ ($R_E$) | $f_{Na}^V$‡ | $\Gamma$ | $\lambda_0$ | $x_0$ | $w_0$ | $v_0$ (m/s) | $a_0$ (m/s) | $r_c$ ($R_E$) | $M_0$ |
|---|---|---|---|---|---|---|---|---|---|
| 2 * | 0.25 | 1.165 | 13.77 | 6.2E-3 | 4.7E-5 | 0.0145 | 1507 | 322.6 | 9E-6 |
| 2 * | 0.20 | 1.147 | 12.61 | 0.047 | 0.034 | 28.4 | 1575 | 42.46 | 0.017 |
| 2 † | 0.15 | 1.132 | 13.37 | 0.049 | 0.025 | 21.1 | 1529 | 40.49 | 0.013 |
| 2 † | 0.10 | 1.125 | 12.19 | 0.081 | 0.106 | 115 | 1601 | 24.72 | 0.067 |
| 3 * | 0.60 | 1.200 | 11.80 | 4.7E-3 | 3E-4 | 0.066 | 1329 | 638.3 | 5E-5 |
| 3 * | 0.50 | 1.180 | 11.16 | 0.046 | 0.075 | 50.7 | 1367 | 65.22 | 0.034 |
| 3 † | 0.60 | 1.188 | 12.63 | 1.7E-4 | 2.3E-9 | 1E-07 | 1285 | 17650 | 7E-11 |
| 3 † | 0.50 | 1.162 | 11.95 | 0.0455 | 0.0461 | 31.0 | 1321 | 65.93 | 0.022 |

The adopted surface density for all the calculated disk models is $\sigma_T = 5 \times 10^7 kg/m^2$

* Atmosphere in equilibrium with metal-saturated magma layer, $T_{eq}$=3,000 K, log $fO_2$ = IW – 2

† Atmosphere in equilibrium with metal-saturated magma layer, $T_{eq}$=3,000 K, log $fO_2$ = IW – 1

‡ The BSE abundances of sodium is taken as 2700 ppm (McDonough and Sun, 1995)